\documentstyle[aps,prb,multicol,epsf]{revtex}

\newcommand{\be}{\begin{equation}}
\newcommand{\ee}{\end{equation}}
\newcommand{\bea}{\begin{eqnarray}}
\newcommand{\eea}{\end{eqnarray}}
\newcommand{\bml}{\begin{mathletters}}
\newcommand{\eml}{\end{mathletters}}
\renewcommand{\Re}{\mathop{\rm Re}\nolimits}
\renewcommand{\Im}{\mathop{\rm Im}\nolimits}
\newcommand{\r}{\rho_{\rm int}}

\newcommand{\lbr}{\left(}
\newcommand{\rbr}{\right)}
\newcommand{\D}{\Delta}
\newcommand{\SN}{\frac{\tau_S}{\tau_N}}
\newcommand{\NS}{\frac{\tau_N}{\tau_S}}
\newcommand{\ds}{\displaystyle}

\begin{document}
\draft

\title{Superconductive properties of thin dirty {\it SN} bilayers}

\author{Ya. V. Fominov and M. V. Feigel'man}

\address{L.~D.~Landau Institute for Theoretical Physics, 117940 Moscow,
Russia}

\date{December 20, 2000}
\maketitle

\begin{abstract}
The theory of superconductivity in thin {\it SN} sandwiches (bilayers) in the
diffusive limit is developed within the standard Usadel equation method, with
particular emphasis on the case of very thin superconductive layers, $d_S
\ll d_N$. The proximity effect in the system is governed by the interlayer
interface resistance (per channel) $\r$. The case of {\it relatively} low
resistance (which can still have large absolute values) can be completely
studied analytically. The theory describing the bilayer in this limit
is of BCS type but with the minigap (in the single-particle density of
states) $E_g \ll \D$ substituting the order parameter $\D$ in the standard
BCS relations; the original relations are thus severely violated. In
the opposite limit of an opaque interface, the behavior of the system is in
many respects close to the BCS predictions. Over the entire range of $\r$, the
properties of the bilayer are found numerically. Finally, it is shown that
the results obtained for the bilayer also apply to more complicated
structures such as {\it SNS} and {\it NSN} trilayers, {\it SNINS} and {\it
NSISN} systems, and {\it SN} superlattices.
\end{abstract}


\begin{multicols}{2}
\narrowtext

% 111111111111111111111111111111111111111111111111111111111111111111111111111
\section{Introduction}

It is well known that the majority of metallic superconductors is
well described by the classical BCS theory of superconductivity.~\cite{BCS}
One of the main qualitative features of the BCS theory is a simple relation
between the superconductive transition temperature $T_c$ and the
low-temperature value of the energy gap for $s$-wave superconductors: $\D(0)
= 1.76\, T_c$. Experimentally, violations of this simple relation are
considered as a sign of some unusual pairing symmetry or even of a non-BCS
pairing mechanism; many theories of unconventional superconductivity were
developed during the last decade, mainly in relation with high-$T_c$
materials.  Recently, an evident example of such a violation of the BCS
theory predictions was found in experiments by Kasumov {\it et
al.},~\cite{Kasumov} who studied current-voltage characteristics of a
carbon nanotube contact between two metallic bilayers (sandwiches) made of
ordinary metals, tantalum and gold. The observed value of the low-temperature
Josephson critical current is 40 times larger than the maximum expected
(Ambegaokar-Baratoff) value~\cite{AB} $I_c = \pi \D (0)/2 eR_{\rm tube}$,
where the energy gap of the bilayer $\D (0)$ is estimated from its transition
temperature. The source of such discrepancy is not clear at present. The most
recent experiments~\cite{Kasumov_new} demonstrate the existence of intrinsic
superconductivity in carbon nanotubes. However, the discrepancy could be due
to unusual superconductive properties of the bilayers. The aim of the present
paper is to investigate these properties.

An essential feature of the experiment~[\onlinecite{Kasumov}] was that the
superconductive layer in the bilayer was very thin ($d_S/d_N$ = 5~nm/100~nm =
1/20). In the present paper, we investigate such a bilayer both analytically
and numerically, calculating quantities characterizing the
superconductivity in this proximity system: the order parameter $\D$, the
density of the superconducting electrons $n$, the critical temperature $T_c$,
the (mini)gap $E_g$ in the single-particle density of states (DOS) and the DOS
$\nu(E)$ itself, the critical magnetic field $H_c$ parallel to the bilayer
and the upper critical field $H_{c2}$ perpendicular to the bilayer. In our
calculations, the parameter controlling the strength of the proximity effect
is the (dimensionless) resistance of the {\it SN} interface per channel $\r$,
which is related to the total interface resistance $R_{\rm int}$ as
\be
\label{rho_int} R_{\rm int} = \frac{R_q \r}{2 N_{\rm ch}},
\ee
where $R_q= h/e^2$ is the quantum resistance, and $N_{\rm ch} = {\cal A}/
(\lambda_F/2)^2$, with $\lambda_F$ being the Fermi wave-length, is the number
of channels in the interface of area ${\cal A}$. We choose $\lambda_F$
referring to the {\it S} layer.

Our results show that the values of the interface resistance can be divided
into three ranges:
(a) at large resistance, many characteristics of the superconductor [$\D$,
$n$, $T_c$, $\nu(E)$, $H_c$, $H_{c2}$] are almost unaffected by the presence
of the normal layer, {\it i.e.}, this is the BCS limit
(however, we note that $E_g$ does not coincide with
the order parameter, and even vanishes as $\r$ increases);
(b) at low resistance, the theory describing the bilayer is of BCS type but
with the order parameter $\D$ substituted by the minigap $E_g$
(for instance, $E_g = 1.76\, T_c$, whereas $E_g \ll \D$); the
original BCS relations are thus severely violated;
(c) at intermediate resistance, the behavior of the system interpolates
between the above two regimes.

The paper is organized as follows. In Section~\ref{sec:usadel}, we formulate
the standard technique of the Usadel equations for dirty
systems,~\cite{Usadel} and introduce a convenient angular parameterization of
the quasiclassical Green functions entering these equations.
In Section~\ref{sec:bilayer}, we apply the Usadel equations to the thin
bilayer that we intend to discuss, and present numerical results for $\D$,
$n$, $T_c$, and $\nu(E)$. We start an analytical analysis of the Usadel
equations for the bilayer by calculating the minigap $E_g$ in the density of
states in the two limiting cases of high and low interface resistance
(Section~\ref{sec:minigap}). Then, in Section~\ref{sec:anderson}, we elucidate
the structure of the theory describing
the system in the so-called Anderson  limit (of relatively low resistance),
finding $\D$, $n$, $T_c$, $E_g$, and $\nu(E)$.
The critical magnetic fields $H_c$ (parallel to the bilayer) and $H_{c2}$
(perpendicular to the bilayer) are calculated in Sections~\ref{sec:H_c}
and~\ref{sec:H_c2}, respectively. In Section~\ref{sec:SNS}, we show that
the results obtained for the bilayer also apply to more complicated structures
such as {\it SNS} and {\it NSN} trilayers, {\it SNINS} and {\it NSISN}
systems, and {\it SN} superlattices. The relationship between our results and
the experiment~[\onlinecite{Kasumov}] that has stimulated our research is
discussed in Section~\ref{sec:discussion}. Finally, we present our conclusions
(Section~\ref{sec:conclusion}).

% 222222222222222222222222222222222222222222222222222222222222222222222222222
\section{Method} \label{sec:usadel}

\subsection{Usadel equation}

Equilibrium properties of dirty systems are described~\cite{LO} by the
quasiclassical
retarded Green function $\hat R({\bf r},E)$, which is a $2\times 2$ matrix in
the Nambu space satisfying the normalization condition $\hat R^2 =\hat 1$.
The retarded Green function obeys the Usadel equation
\be \label{usadel}
D \nabla (\hat R \nabla \hat R) + i[\hat H,\,\hat R] =0.
\ee
Here the square brackets denote the commutator,
$D ={\rm v} l/3$ is the diffusion constant with ${\rm v}$ and $l$ being the
Fermi velocity and the elastic mean free path,
$\hat H = E \hat\sigma_z + \hat\D({\bf r})$ with $E$ being the energy,
whereas $\hat\sigma_z$ (the Pauli matrix) and $\hat\D({\bf r})$ are given by
\be
\hat\sigma_z = \lbr \begin{array}{cc} 1 & 0\\ 0 & -1 \end{array} \rbr,\qquad
\hat\D = \lbr \begin{array}{cc} 0 & \D\\ -\D^* & 0 \end{array} \rbr.
\ee
The order parameter $\D({\bf r})$ must be determined self-consistently from
the equation
\be
\hat\D({\bf r}) = \frac{\nu_0 \lambda}4 \!\int_0^{\omega_D} \!\!\! dE\,
\tanh \!\lbr\! \frac E{2T} \!\rbr\!
\left[ \hat R({\bf r},E) - \hat R({\bf r},-E)
\right]_{\rm o.d.} ,
\ee
where the subscript $\rm o.d.$ denotes the off-diagonal part,
$\nu_0 = m^2 {\rm v} /\pi^2$ is the normal-metal density of states at the
Fermi level, $\lambda$ is the effective constant of electron-electron
interaction in the {\it S} layer (whereas we assume $\lambda=0$ and hence $\D
=0$ in the {\it N} layer), and integration is cut off at the Debye energy
$\omega_D$ of the {\it S} material.

Equation~(\ref{usadel}) should be supplemented with the appropriate boundary
conditions at an interface, which read~\cite{KL}
\be \label{interface}
\sigma_l \lbr \hat R_l \nabla_{\bf n} \hat R_l \rbr =
\sigma_r \lbr \hat R_r \nabla_{\bf n} \hat R_r \rbr =
\frac{g_{\rm int}}2 \left[ \hat R_l,\,\hat R_r \right],
\ee
where the subscripts $l$ and $r$ designate the left and right electrode,
respectively; $\sigma$ is the conductivity of a metal in the normal state, and
$g_{\rm int}=G_{\rm int}/ {\cal A}$ (with $G_{\rm int} = 1/ R_{\rm int}$) is
the conductance of the interface per unit area when both left and right
electrodes are in the normal state. $\nabla_{\bf n}$ denotes the projection of
the gradient upon the unit vector ${\bf n}$ normal to the interface.

The system of units in which the Planck constant and the speed of light equal
unity ($\hbar=c=1$) is used throughout the paper.

\subsection{Angular parameterization of the Green function}
\label{subsec:theta}

The normalization condition allows the angular parameterization of the
retarded Green function:
\be
\hat R = \lbr \begin{array}{cc} \cos\theta & -i e^{i\varphi}
\sin\theta\\ i e^{-i\varphi}\sin\theta & -\cos\theta \end{array} \rbr,
\ee
where $\theta=\theta({\bf r},E)$ is a complex angle which characterizes the
pairing, and $\varphi=\varphi({\bf r},E)$ is the real superconducting phase.
The off-diagonal elements of the matrix $\hat R$ describe~\cite{LO} the
superconductive correlations, vanishing in the bulk of a normal metal ($\theta
=0$).

The Usadel equation takes the form
\bml \label{usadel_theta}
\be \label{usadel_theta_1}
\frac D2 \nabla^2\theta + \left[ iE - \frac D2 \lbr \nabla\varphi \rbr^2
\cos\theta \right] \sin\theta + \left| \D \right| \cos\theta = 0,
\ee
\be \label{usadel_theta_2}
\nabla \lbr \sin^2 \theta\, \nabla\varphi \rbr = 0.
\ee
\eml
The corresponding boundary conditions are
\bml
\label{b}
\be
\sigma_l \nabla_{\bf n} \theta_l = g_{\rm int} \left[ \cos (\varphi_r
-\varphi_l ) \cos\theta_l \sin\theta_r - \sin\theta_l \cos\theta_r \right],
\ee
\be
\sigma_r \nabla_{\bf n} \theta_r = g_{\rm int} \left[ \cos\theta_l
\sin\theta_r - \cos (\varphi_r -\varphi_l )\sin\theta_l \cos\theta_r \right],
\ee
\bea
\sigma_l \sin^2 \theta_l\, \nabla_{\bf n} \varphi_l &=& \sigma_r \sin^2
\theta_r\, \nabla_{\bf n} \varphi_r \nonumber \\
&=& g_{\rm int} \sin (\varphi_r -\varphi_l) \sin\theta_l \sin\theta_r.
\label{b_3}
\eea
\eml

The self-consistency equation for the order parameter $\D({\bf r})$ takes the
form
\be \label{Delta}
\D =\frac{\nu_0 \lambda}2 \int_0^{\omega_D} dE\,\tanh\!\lbr\!\frac E{2T} \!
\rbr \Im\left[\sin\theta\right] e^{i\varphi}.
\ee

The above equations are written in the absence of an external magnetic field.
To take account of the magnetic field, it is sufficient to substitute the
superconducting phase gradient in the Usadel equations~(\ref{usadel_theta})
by its gauge invariant form $2m {\bf v} =\nabla\varphi+2e{\bf A}$, where
${\bf A}$ is the vector potential and ${\bf v}$ denotes the supercurrent
velocity.

Physical properties of the system can be expressed in terms of the
pairing angle $\theta({\bf r},E)$.
The single-particle density of states $\nu({\bf r},E)$ and the density of the
superconducting electrons $n({\bf r})$ are given by
\bea
\label{nu}
\nu &=& \nu_0 \Re\,[\cos\theta],\\
\label{n}
n &=& \frac{2m\sigma}{e^2} \int_0^\infty dE\,\tanh\!\lbr\!\frac E{2T} \!\rbr
\Im \left[\sin^2\theta\right],
\eea
where $m$ and $e$ are the electron's mass and the absolute value of its charge.
The total number of single-particle states in a metal is the same in the
superconducting and normal states, which is expressed by the constraint
\be
\int_0^\infty dE \left[ \nu({\bf r},E) -\nu_0 \right] =0.
\ee

\subsection{Simple example: the BCS case}

The simplest illustration for the above technique is the BCS case, when the
order parameter $\D({\bf r})=\D_{BCS}$ is spatially constant. Its phase can be
set equal to zero, $\varphi=0$. Then the Usadel equations~(\ref{usadel_theta})
are trivially solved, and we can write the answer in terms of the sine and the
cosine of the pairing angle:
\bml \label{sin_cos_BCS}
\bea
\sin\theta_{BCS} (E) &=& \frac{i\Delta_{BCS}}{\sqrt{E^2-\Delta_{BCS}^2}},\\
\label{cos_BCS}
\cos\theta_{BCS} (E) &=& \frac E{\sqrt{E^2-\Delta_{BCS}^2}}.
\eea
\eml
An infinitesimal term $i0$ should be added to the energy $E$ to take the
retarded nature of the Green function $\hat R$ into account, which yields
\be \label{Im_sin_sq_BCS}
\Im \left[\sin^2 \theta_{BCS} (E) \right] =\frac \pi 2 \Delta_{BCS}\, \delta
(E-\Delta_{BCS}).
\ee

The usual BCS relations are straightforwardly obtained from
Eqs.~(\ref{Delta}), (\ref{nu}), (\ref{n}) (for simplicity, we consider
the case of zero temperature):
\bea
\D_{BCS} &=& 2\omega_D \exp\lbr -\frac 2{\nu_0 \lambda} \rbr,\\
\label{nu_BCS}
\nu_{BCS}(E) &=& \left\{ \begin{array}{ll} \ds 0, &\;\mbox{if } E <\D_{BCS}\\
\ds \nu_0 \frac E{\sqrt{E^2- \D_{BCS}^2}}, &\;\mbox{if } E >\D_{BCS}
\end{array} \right. ,\\
\label{n_BCS}
n_{BCS} &=& \pi \frac{m\sigma}{e^2} \D_{BCS}.
\eea
The critical temperature must be determined from Eq.~(\ref{Delta}) with
vanishing $\Delta (T_c)$; the result is
\be
\label{T_c_BCS}
\D_{BCS} (0) = \frac\pi\gamma T_c^{BCS},
\ee
where $\gamma \approx 1.78$ is Euler's constant.

% 333333333333333333333333333333333333333333333333333333333333333333333333333
\section{Usadel equations for a thin bilayer} \label{sec:bilayer}

Let us consider a {\it SN} bilayer consisting of a normal metal ($-d_N<z<0$)
in contact (at $z=0$) with a superconductor ($0<z<d_S$). We assume that the
layers are thin (this assumption will be discussed in
Sec.~\ref{sec:discussion}) and can be regarded as uniform, which allows us to
set the order parameter $\D$ equal to a constant in the superconductive layer
(we choose its phase $\varphi$ equal to zero). At the same time, we suppose
that electron-electron interaction is absent in the normal layer: $\lambda=0$,
hence $\D=0$, although the superconductive correlations ($\theta \neq 0$)
exist in the {\it N} layer due to the
proximity effect. The Usadel equations~(\ref{usadel_theta}) take the form
\bml \label{usadel_bi}
\bea
\frac{D_N}2 \frac{\partial^2\theta_N}{\partial z^2} +iE\sin\theta_N &=& 0,\\
\frac{D_S}2 \frac{\partial^2\theta_S}{\partial z^2} +iE\sin\theta_S+
\D\cos\theta_S &=& 0,
\eea
\eml
where $\theta_N$ and $\theta_S$ denote the pairing angle $\theta$ at $z<0$
and $z>0$, respectively.

The boundary conditions~(\ref{b}) reduce to
\be \label{gran_uslovija}
\sigma_N \frac{\partial\theta_N}{\partial z} =
\sigma_S \frac{\partial\theta_S}{\partial z} =
g_{\rm int} \sin(\theta_S-\theta_N).
\ee

Equations~(\ref{usadel_bi}) can be integrated once, yielding
\bml \label{integrated}
\bea
\frac{D_N}4 \lbr \frac{\partial\theta_N}{\partial z} \rbr^2 -iE\cos\theta_N
&=& f_N,\\
\frac{D_S}4 \lbr \frac{\partial\theta_S}{\partial z} \rbr^2 -iE\cos\theta_S +
\D\sin\theta_S &=& f_S.
\eea
\eml
The functions $f_N(E)$ and $f_S(E)$ are determined from the boundary
condition $\partial\theta/\partial z=0$ at the nontransparent outer surfaces
of the bilayer, which give
\bea
f_N(E) &=& -iE\cos\theta_N(-d_N,E), \nonumber \\
f_S(E) &=& -iE\cos\theta_S(d_S,E)+\D\sin\theta_S(d_S,E).
\eea
Let us denote $\theta_N(E)=\theta_N(-d_N,E)$, $\theta_S(E) =\theta_S(d_S,E)$.
Because of the uniformity of the layers, the functions
$\theta_N(z,E)$ and $\theta_S(z,E)$ are nearly spatially constant. However,
in order to determine them, we should take account of their weak spatial
dependence and make use of the boundary conditions at the {\it SN} interface.
Substituting
\bea
\theta_N(z,E) &=& \theta_N(E)+\delta\theta_N(z,E), \nonumber \\
\theta_S(z,E) &=& \theta_S(E)+\delta\theta_S(z,E)
\eea
into Eqs.~(\ref{integrated}) and linearizing them with respect to $|\delta
\theta_N(z,E)|$, $|\delta\theta_S(z,E)| \ll 1$, we find the solution.
Finally, boundary conditions at the {\it SN} interface lead to
\bea \label{main}
-i\tau_N E\sin\theta_N(E) &=& i\tau_S E\sin\theta_S(E) +
\tau_S\D\cos\theta_S(E) \nonumber\\
&=&\sin\left[ \theta_S(E)-\theta_N(E) \right],
\eea
where we have denoted $\tau_N = 2\sigma_N d_N /D_N\, g_{\rm int}$, $\tau_S =
2\sigma_S d_S /D_S\, g_{\rm int}$. Using the definition of the interface
resistance per channel~(\ref{rho_int}), we can represent these quantities as
\be
\tau_N = 2\pi \frac{{\rm v}_N d_N}{{\rm v}_S^2} \r,\qquad
\tau_S = 2\pi \frac{d_S}{{\rm v}_S} \r,
\ee
with ${\rm v}_N$ and ${\rm v}_S$ being the Fermi velocities in the {\it N} and
{\it S} layers.
The ratio $\tau_N /\tau_S = {\rm v}_N d_N /{\rm v}_S d_S$,
which is independent of interface properties, can also be
interpreted as the ratio of the {\it global} densities of states (per
energy interval) in the two layers considered,
\be
\frac{\tau_N}{\tau_S} = \frac{{\cal A} d_N \nu_{0N}}{{\cal A} d_S \nu_{0S}}.
\ee
The latter interpretation will prove useful for further analysis.

Having solved the boundary conditions~(\ref{main}), we can
determine all equilibrium properties of the system (because knowledge of
$\theta_S$, $\theta_N$ implies knowledge of the retarded Green function
$\hat R$).

A useful representation of the boundary conditions~(\ref{main}) is obtained as
follows. Excluding $\theta_N(E)$ from Eq.~(\ref{main}), we arrive at a single
equation for the function $\theta_S(E)$, which can be written, in terms of
$Z=\exp(i\theta_S)$, as a polynomial equation
\bea
\label{main_Z}
&& iC_6 Z^6 + C_5 Z^5 + iC_4 Z^4 + C_3 Z^3 + iC_2 Z^2 + C_1 Z +iC_0 \nonumber \\
&& = 0
\eea
with real coefficients
\bea
C_6 &=& -\tau_N E \lbr\SN\rbr^2 \left[ 1+\frac\D E \right]^2, \nonumber \\
C_5 &=& \left[ 1-\lbr \tau_N E\rbr^2 \right] \lbr\SN\rbr^2 \left[ 1+\frac\D E
\right]^2 -1, \nonumber \\
C_4 &=& -\tau_N E \lbr\SN\rbr^2 \left[ 3\lbr \frac\D E \rbr^2+ 2\frac{\D}{E}
-1 \right], \nonumber \\
C_3 &=& 2-2\left[ 1-\lbr \tau_N E \rbr^2 \right] \lbr\SN\rbr^2 \left[ 1-
\lbr \frac{\D}E \rbr^2 \right], \nonumber \\
C_2 &=& -\tau_N E \lbr\SN\rbr^2 \left[ 3\lbr \frac{\D}E \rbr^2 -2\frac\D E -1
\right], \nonumber \\
C_1 &=& \left[ 1-\lbr \tau_N E\rbr^2 \right] \lbr\SN\rbr^2 \left[ 1-\frac\D E
\right]^2 -1, \nonumber \\
C_0 &=& -\tau_N E \lbr\SN\rbr^2 \left[ 1-\frac\D E \right]^2.
\label{coeff}
\eea

During further analysis, the choice between the boundary conditions in the
forms~(\ref{main}) and~(\ref{main_Z}) will be a matter of convenience.

\subsection{Critical temperature}

The critical temperature of the bilayer $T_c$ is defined from the condition
of vanishing of the order parameter $\D$. Near $T_c$, the superconducting
correlations are very small, $|\theta|\ll 1$; nevertheless, the
self-consistency equation~(\ref{Delta}) has a nonzero solution $\D \neq 0$.

Linearizing the boundary conditions~(\ref{main}) with respect to $\theta_N$
and $\theta_S$, we readily find the solution:
\be
\theta_S(E) = i\frac\D E \lbr 1- \frac{\tau_N}{\tau_S+\tau_N-i\tau_S\tau_N E}
\rbr.
\ee
Substituting this into the self-consistency equation~(\ref{Delta}) and
simplifying both its sides by $\D$, we obtain an equation determining $T_c$,
which can be cast into the form
\bea
\ln\frac{T_c^{BCS}}{T_c} = \frac{\tau_N}{\tau_S+\tau_N}
\left[ \vphantom{\sqrt{\lbr\frac{\tau_S+\tau_N}{\tau_S\tau_N\omega_D}\rbr^2}}
\right.
&&\psi\lbr \frac 12 +\frac{\tau_S +\tau_N}{2\pi T_c \tau_S \tau_N} \rbr
-\psi \lbr \frac 12 \rbr \nonumber \\
&&\left.
-\ln\sqrt{1+\lbr\frac{\tau_S+\tau_N}{\tau_S\tau_N\omega_D} \rbr^2} \right],
\label{T_c}
\eea
where $\psi(x)$ denotes the digamma function.
A similar formula (except the logarithmic term in the r.h.s.) was obtained by
McMillan~\cite{McMillan,Gamma,T_c_McMillan} (see also
Ref.~[\onlinecite{Golubov}]). The logarithmic term in the r.h.s. takes account
of the finiteness of the Debye energy $\omega_D$; it becomes important only in
the limit of a perfect interface (the Cooper limit), {\it i.e.}, when
$\tau_S\tau_N\omega_D /(\tau_S+\tau_N) \ll 1$.
Equation~(\ref{T_c}) can be solved numerically over the entire range of $\r$
(see Sec.~\ref{sec:numerical}); the analytical solution can be found in
limiting cases (see Sec.~\ref{sec:anderson}).

\subsection{Numerical results} \label{sec:numerical}

The solution of Eq.~(\ref{main_Z}) can be found numerically. To this end, we
solve the system of two nonlinear equations for the functions $\Re Z(E)$ and
$\Im Z(E)$, using the modified Newton method with normalization.

The solution depends on the bilayer's parameters: the thicknesses of the
layers, characteristics of materials constituting the bilayer, and the
quality of the {\it SN} interface. This dependence enters Eqs.~(\ref{main_Z}),
(\ref{coeff}) via $\tau_N$ and $\tau_S$.
For numerical calculations, we assume the characteristics of the bilayer to
be the same as in the experiment by Kasumov {\it et al.}~\cite{Kasumov}
The superconductive layer is made of tantalum, $d_S=5$~nm, and the normal
layer is made of gold, $d_N=100$~nm. Approximate experimental values of the
conductivities are~\cite{Kasumov_priv} $\sigma_S = 0.01\, \mu\Omega^{-1}\,
{\rm cm}^{-1}$ and $\sigma_N = 1\, \mu\Omega^{-1}\, {\rm cm}^{-1}$. In order
to calculate the Fermi characteristics of tantalum and gold, we use the values
of the Fermi energy $E_F ({\rm Ta}) = 11$~eV, $E_F ({\rm Au}) = 5.5$~eV,
and the free electrons model.~\cite{free_el}

% -----------------------------   FIGURE   ----------------------------------
\begin{figure}[p]
%\centerline{\epsfxsize=8.7cm \epsfbox{wide_rho.eps}}
\centerline{\epsfxsize=\hsize \epsfbox{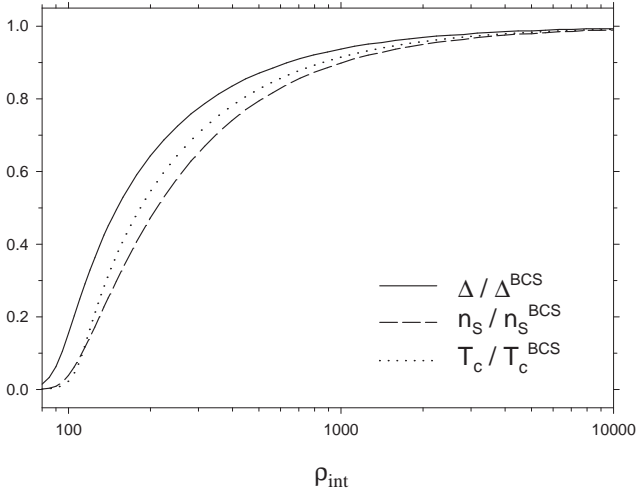}} \vspace{2mm}
\caption{Dependence of the order parameter in the {\it S} layer $\D$, of the
superconducting electrons' density in the {\it S} layer $n_S$, and of the
bilayer's critical temperature $T_c$ on the interface resistance per channel
$\r$ at zero temperature. All the quantities are normalized by the
corresponding BCS values. The discrepancy between the curves implies a
violation of the BCS relations between $\D$, $n_S$, and $T_c$.}
\label{fig:wide_rho}
\end{figure}
% -----------------------------   FIGURE   ----------------------------------
\begin{figure}
%\centerline{\epsfxsize=8.8cm \epsfbox{narr_rho.eps}}
\centerline{\epsfxsize=\hsize \epsfbox{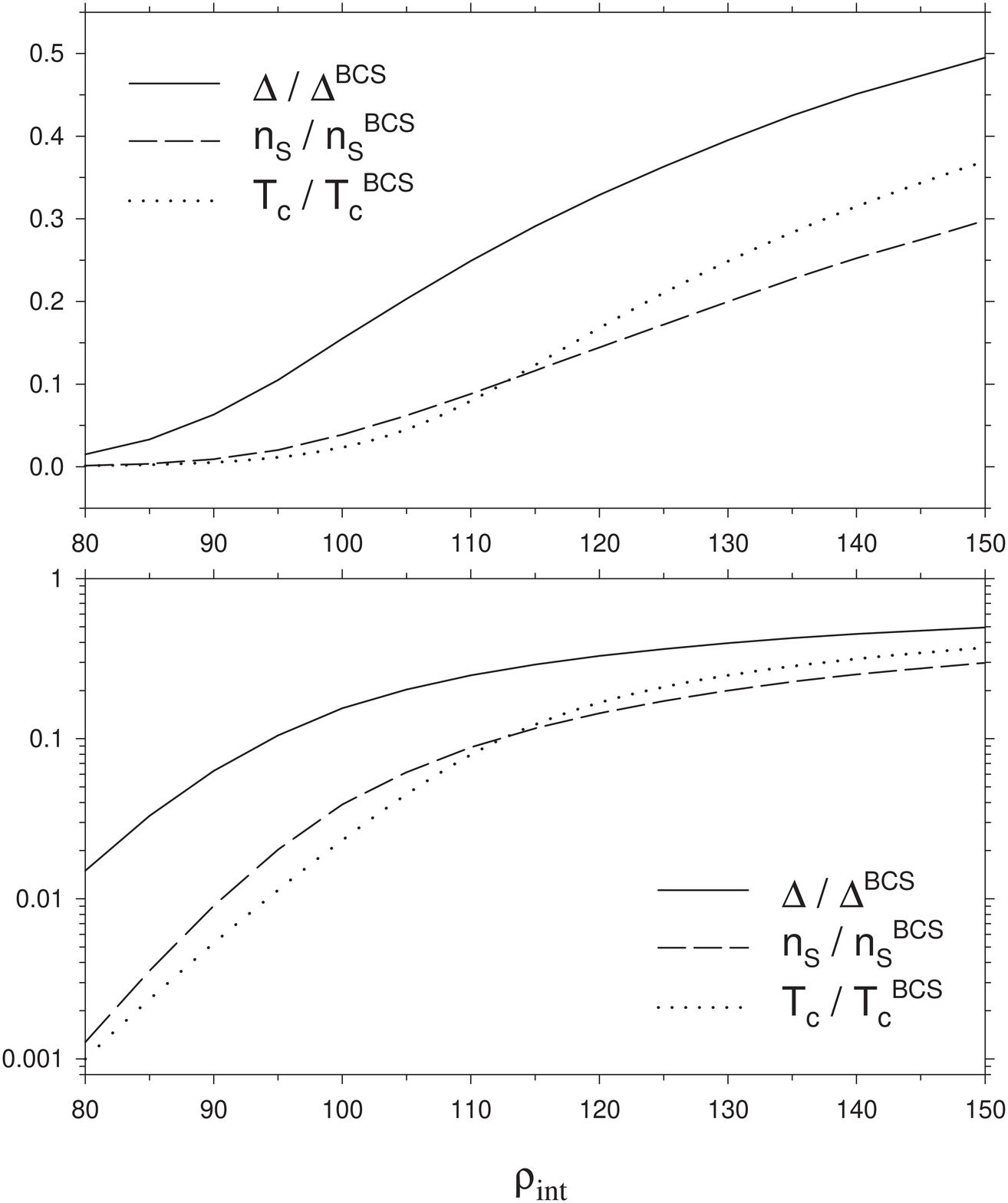}} \vspace{2mm}
\caption{Zoomed part of Fig.~\protect\ref{fig:wide_rho}. In the shown range
of relatively small resistance $\r$, the BCS relations between $\D$, $n_S$,
and $T_c$ are severely violated. The upper and lower graphs differ only in the
scaling of the ordinate axis (normal and logarithmic, respectively).}
\label{fig:narr_rho}
\end{figure}
% -----------------------------   FIGURE   ----------------------------------
\begin{figure}
%\centerline{\epsfxsize=8.8cm \epsfbox{t_depend.eps}}
\centerline{\epsfxsize=\hsize \epsfbox{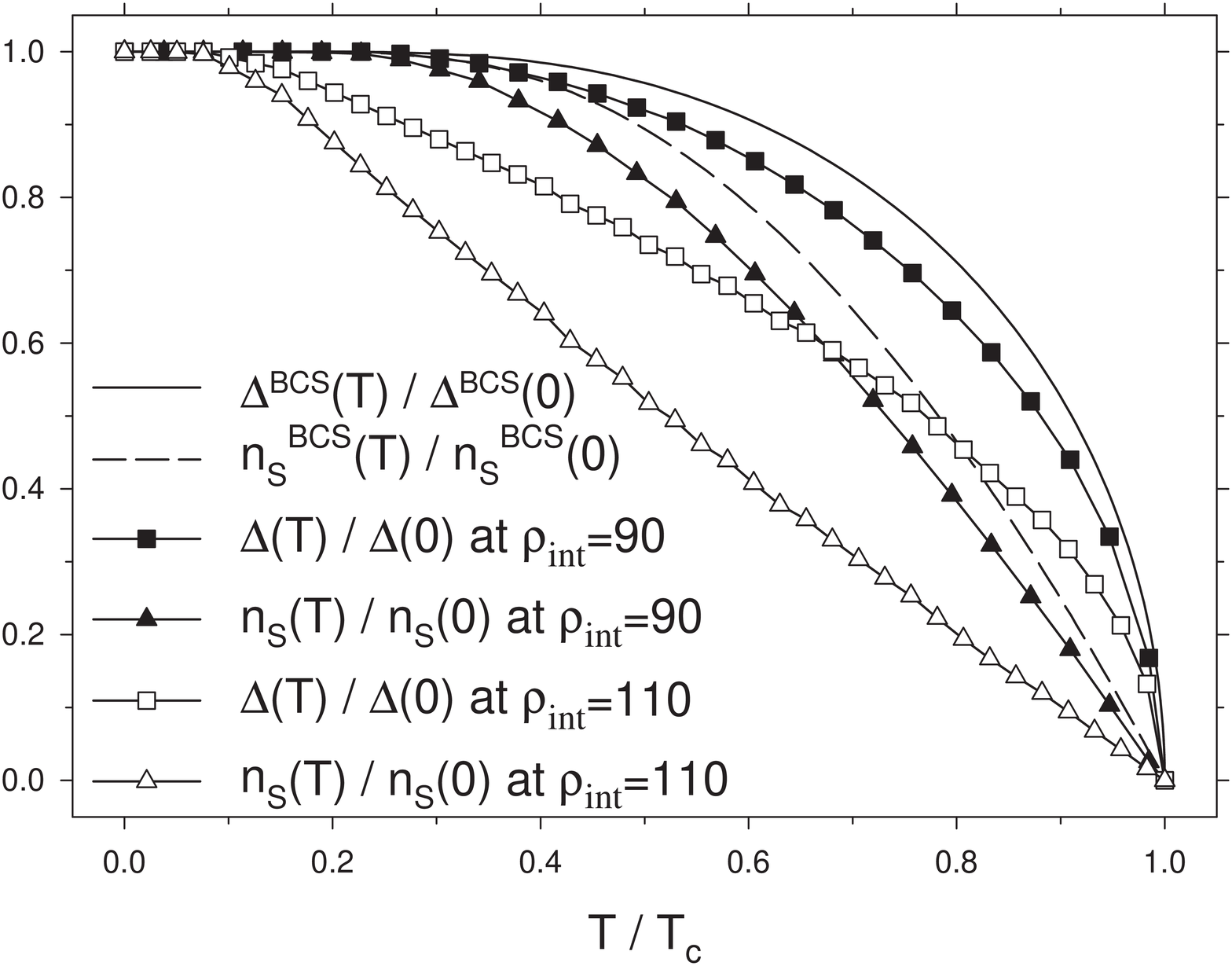}} \vspace{1mm}
\caption{Temperature dependence of $\D$ and $n_S$ at $\r =$ 90 and 110.
The temperature is normalized by the critical value $T_c$, which depends on $\r$;
$\D$ and $n_S$ are normalized by their zero-temperature values.
For comparison, the same dependence is also plotted for the BCS case.}
\label{fig:t_depend}
\end{figure}
\vspace{-8mm}
% -----------------------------   FIGURE   ----------------------------------
\begin{figure}
%\centerline{\epsfxsize=8.5cm \epsfbox{dos.eps}}
\centerline{\epsfxsize=\hsize \epsfbox{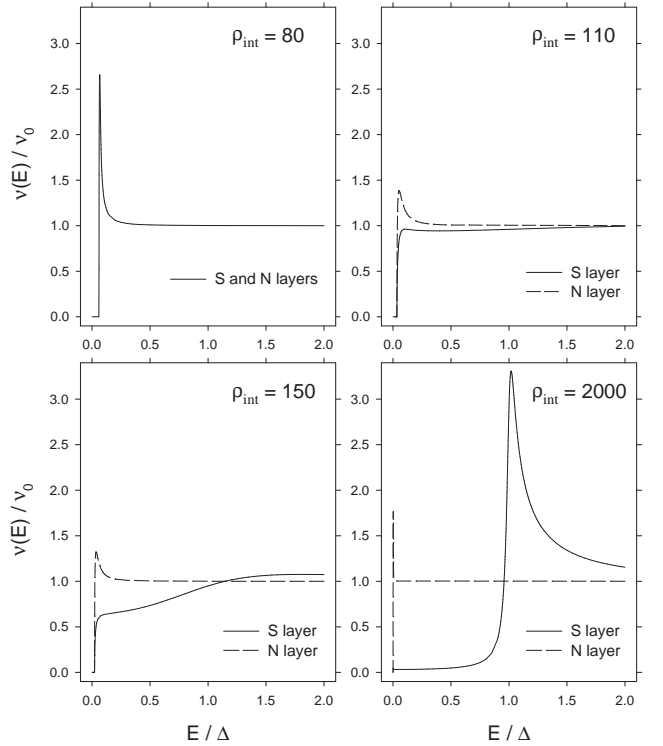}} \vspace{1mm}
\caption{Energy dependence of the single-particle DOS, normalized by the
normal-metal DOS, $\nu(E)/\nu_0$, in the {\it S} and {\it N} layers at $\r=$
80, 110, 150, and 2000. The energy $E$ is normalized by the order parameter
$\D$, which is different in all the four cases. [To avoid confusion, we note
that actual (in absolute units) relation between the minigaps in the DOS does
not correspond to what is seen from the Figure.] The Figure demonstrates
drastic difference of the DOS in the bilayer from the BCS case
[Eq.~(\ref{nu_BCS})].}
\label{fig:dos}
\end{figure}

Once the parameters have been specified, the solution of Eq.~(\ref{main_Z})
depends only on the interface resistance $\r$. Having found the function
$Z(E)$ [which is equivalent to finding $\theta_S(E)$], we start from the case
of zero temperature, $T=0$, and study the dependence of the order parameter
$\D$ and of the superconducting electrons' density in the {\it S} layer $n_S$
[Eqs.~(\ref{Delta}), (\ref{n})] on $\r$. The results are plotted in
Fig.~\ref{fig:wide_rho}, where we also show the dependence of the critical
temperature $T_c$, determined from Eq.~(\ref{T_c}), on $\r$.

The suppression of $\D$, $n_S$, and $T_c$, in comparison to their BCS values
in the {\it S} layer, is a natural consequence of proximity to the normal
metal. At the same time, there is a possibility of BCS-like behavior, which
implies the BCS relations between the suppressed quantities and the
coincidence of the three curves plotted in Fig.~\ref{fig:wide_rho}. However,
the curves split, and the difference between them is largest for relatively
small values of $\r$. Figure~\ref{fig:narr_rho} presents the range
$80 < \r < 150$ on a larger scale.

Figure~\ref{fig:t_depend} shows the temperature dependence of the order
parameter $\D$ and of the superconducting electrons' density in the {\it S}
layer $n_S$. Although the smaller $\r$ the further it is from the BCS limit
(corresponding to $\r\to\infty$), we observe that at $\r =90$ the curves are
closer to the BCS behavior than at $\r =110$. An explanation of this feature
is given in Sec.~\ref{sec:anderson}.

Finally, the energy dependence of the single-particle density of states in the
{\it S} and {\it N} layers $\nu_{S,N}(E)$ is plotted in Fig.~\ref{fig:dos}.
The density of states in the bilayer is qualitatively different from the BCS
result~(\ref{nu_BCS}). In particular, there is a minigap $E_g$ in the density
of states at energies much smaller then $\D_{BCS}$, and even much smaller then
$\D$ in the bilayer. In the next Section, we find this minigap analytically in
the two limiting cases of small and large interface resistance $\r$.
Another feature which can be seen from Fig.~\ref{fig:dos} is that the order
parameter $\D$ plays the role of a characteristic energy scale of the system
only in the limit of large $\r$ (see the case $\r =2000$). Otherwise, no
peculiarity in the DOS is observed at $E =\D$. Some other aspects of the DOS
behavior will be discussed in Secs.~\ref{sec:minigap}, \ref{sec:anderson}.

% 444444444444444444444444444444444444444444444444444444444444444444444444444
\section{Minigap in the density of states} \label{sec:minigap}

In principle, the energy dependence of the single-particle density of states
is different in the {\it S} and {\it N} layers. At the same time, the gap in
the DOS is a property of the bilayer as a whole; the gap is spatially
independent because there is no localization in the system and each
electronic state extends over the whole bilayer.

The presence of the gap thus implies that the density of states vanishes in
both layers, when the energy is below the gap:
\be
\Re\,[\cos\theta_S] =\Re\,[\cos\theta_N] =0,
\ee
leading to $\theta_S = \frac{\pi}2 +i\vartheta_S$, $\theta_N =
\frac{\pi}2 +i\vartheta_N$,
with real $\vartheta_S$ and $\vartheta_N$. In this case, Eqs.~(\ref{main})
can be written as
\bml \label{two}
\bea
\tanh\vartheta_N &=& \frac{\sinh\vartheta_S + \tau_N E}{\cosh\vartheta_S},\\
\cosh\vartheta_N &=& -\SN \cosh\vartheta_S + \frac{\tau_S \D}{\tau_N E}
\sinh\vartheta_S.
\eea
\eml

Assuming that $\sinh\vartheta_S \gg\tau_N E$ at small energies,
from Eqs.~(\ref{two}) we obtain $\vartheta_N =\vartheta_S$, and, finally,
\be \label{cos}
\cos\theta_S = \cos\theta_N = \frac E{\sqrt{E^2 -E_g^2}},
\ee
with $E_g =\tau_S \D/(\tau_S+\tau_N)$. This is a BCS-like result
[cf.~Eq.~(\ref{cos_BCS})],
although the order parameter $\D_{BCS}$ is substituted by the minigap $E_g$.
The assumption is readily checked, and we obtain
\be \label{small}
E_g = \frac{\tau_S}{\tau_N+\tau_S} \D,\qquad\mbox{if}\,\,\,
\frac{\tau_S \tau_N \D}{\tau_S+\tau_N} \ll 1.
\ee

Now we proceed to the opposite limit of large interface resistance.
Assuming $\sinh\vartheta_S \ll\tau_N E$ and $\sinh\vartheta_S \ll 1$, we
solve Eqs.~(\ref{two}) and finally obtain
\bml
\label{cos_SN}
\bea
\cos\theta_S &=& \frac E{\tau_S\D \sqrt{E^2-1/\tau_N^2}} -i\frac E\D,\\
\cos\theta_N &=& \frac E{\sqrt{E^2-1/\tau_N^2}}.
\eea
\eml
The assumption is readily checked, and the result is
\be \label{large}
E_g = \frac 1{\tau_N},\qquad\mbox{if}\,\,\,
\frac{\tau_S \tau_N \D}{\tau_S+\tau_N} \gg 1.
\ee

\begin{figure}
%\centerline{\epsfxsize=8.6cm \epsfbox{minigap.eps}}
\centerline{\epsfxsize=\hsize \epsfbox{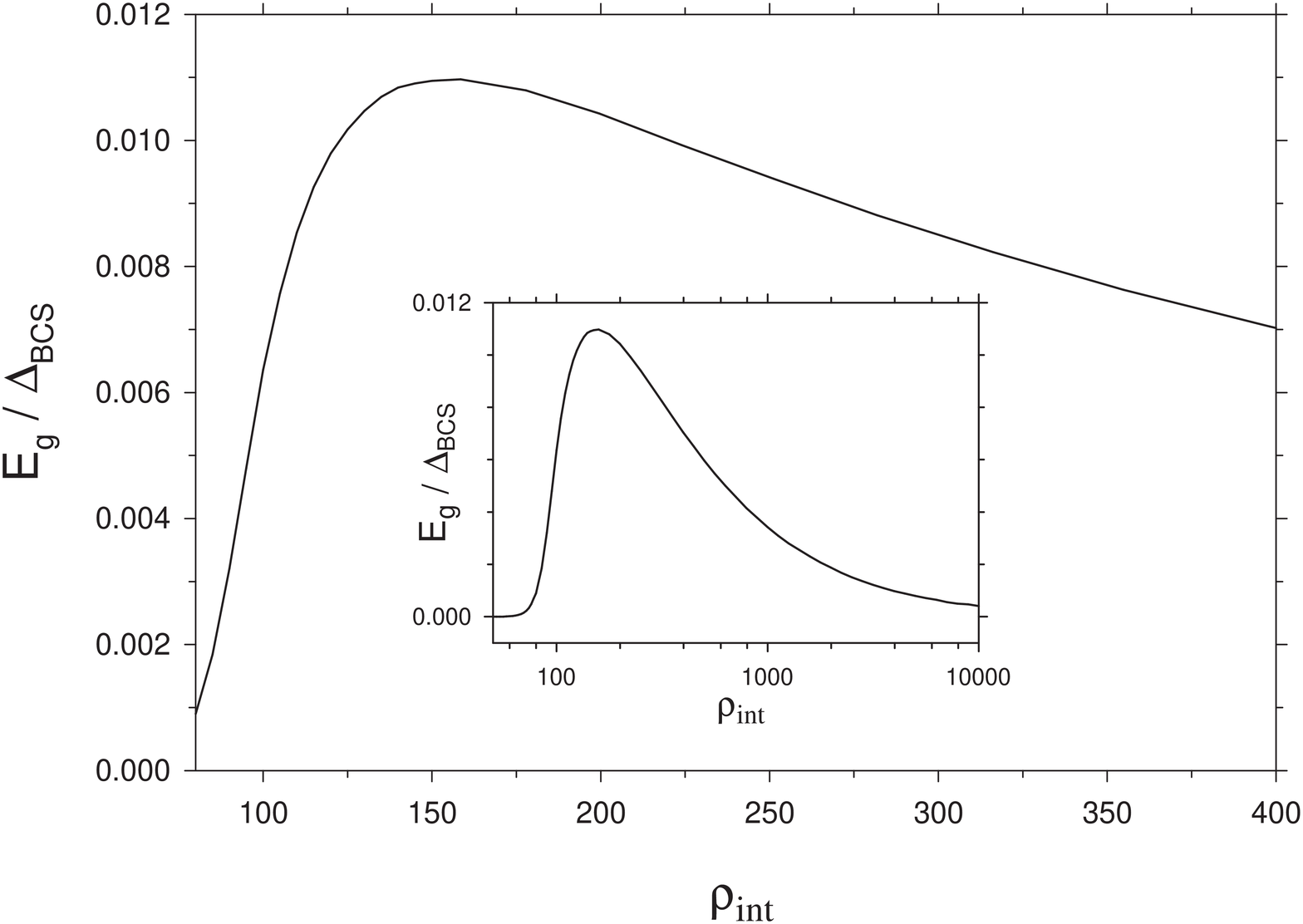}} \vspace{2mm}
\caption{Minigap in the single-particle density of states $E_g$ versus
$\r$. The minigap is normalized by the BCS gap value $\D_{BCS}$. $E_g$ is a
nonmonotonic function of $\r$, reaching its maximum at $\r = 160$. The
inset shows $E_g (\r)$ on a wider (logarithmic) scale over $\r$.}
\label{fig:minigap}
\end{figure}

Equations~(\ref{small}) and~(\ref{large}) imply that $E_g$ is a nonmonotonic
function of the interface resistance: with increase of $\r$, it first
increases at small $\r$ [Eq.~(\ref{small})] and then decreases at
large $\r$ [Eq.~(\ref{large})]. Therefore, $E_g$ reaches its
maximum at some intermediate value of $\r$, corresponding to $\tau_S \tau_N \D /
(\tau_S+\tau_N) \sim 1$, hence $\r\sim 140$.
Numerical results for $E_g$ are shown in Fig.~\ref{fig:minigap}.

At first sight, vanishing of the minigap $E_g$ in the limit of an opaque
interface seems to contradict the general tendency to the BCS behavior.
However, this contradiction is more apparent than real. Actually, the DOS
curve for the {\it S} layer does approach the BCS result~(\ref{nu_BCS}) in
this limit, showing the standard peculiarity at $E\approx \D_{BCS}$. At the
same time, below $\D_{BCS}$, the DOS curve sharply drops to very small values
(which are still finite in contrast to the BCS case), and turns to zero only
at $E = E_g$. Simultaneously, the DOS in the {\it N} layer approaches the
(constant) normal-metal value. The tendency to such behavior is illustrated by
Fig.~\ref{fig:dos}, the case $\r=2000$.
% -----------------------------   FIGURE   ----------------------------------

The results~(\ref{cos}), (\ref{cos_SN}) are valid not only below the minigap
but also right above it (in which case the real parts of $\theta_S$ and
$\theta_N$ differ from $\pi/2$), providing a comparison between the DOS in the
{\it S} and {\it N} layers. Equation~(\ref{cos}) demonstrates the equality of
the DOS in the two layers at relatively small $\r$ (see Fig.~\ref{fig:dos},
the case $\r=80$). Proceeding to the limit of an opaque interface, we should
note that the approximation which lead to Eqs.~(\ref{cos_SN}) fails in a
narrow vicinity of $E_g$ [this fact does not affect the result for the
minigap~(\ref{large}) itself, but the incorrect divergence of the DOS at
$E=E_g$ disappears]. Equations~(\ref{cos_SN}) demonstrate that outside this
region, at energies right above the minigap, the DOS in the {\it N} layer
exceeds the DOS in the {\it S} layer by the large factor $\tau_S \D$
(see Fig.~\ref{fig:dos}, the case $\r=2000$).

Finally, we note that the results of the present Section are similar to those
obtained by McMillan~\cite{McMillan,Gamma} (see also
Ref.~[\onlinecite{Golubov}]).

% 555555555555555555555555555555555555555555555555555555555555555555555555555
\section{Anderson limit} \label{sec:anderson}

In the limit of relatively low interface resistance (the Anderson limit), the
theory describing the bilayer can be developed analytically. The condition
defining this limit is $\tau_S \D$, $\tau_N \D \ll 1$.

First of all, we need to determine $\theta (E)$ [or $Z(E)$] solving
Eq.~(\ref{main}) [or Eq.~(\ref{main_Z})] over the entire range of energies $E$.

In the region $E>\D$, the solution of Eq.~(\ref{main_Z}) can be written as
$Z=1+\delta Z$, with $|\delta Z| \ll 1$. Keeping terms up to the
second order in $\delta Z$, we obtain
\be \label{more_1}
\delta Z = -\frac{\D \lbr 1-i\tau_N E \rbr}{E \lbr
\frac{\tau_S+\tau_N}{\tau_S} -i\tau_N E \rbr}.
\ee
This result is general in the sense that it is valid for arbitrary values of
$\r$.

At $E<\D$, the same calculation as for the minigap leads to the result
\be \label{less_1}
\sin\theta_S = \sin\theta_N = \frac {iE_g}{\sqrt{ E^2 -E_g^2}},
\ee
with the minigap $E_g$ given by Eq.~(\ref{small}).
[To avoid confusion, we note that under the less strict limitations for
$\tau_S \D$, $\tau_N \D$ used in Eq.~(\ref{small}), the BCS-like
results~(\ref{cos})
and~(\ref{less_1}) are valid only up to energies of the order of $E_g$.]

Now $\Im\,[\sin\theta_S]$ is readily calculated, and in the case of zero
temperature, $T=0$, the self-consistency equation~(\ref{Delta}) can be solved,
yielding
\be \label{Delta_anderson}
\frac{\D}{\D_{BCS}}= \lbr \frac{\tau_S+\tau_N}{\tau_S} \rbr \left[
\frac{\D_{BCS}}{2\omega_D}
\sqrt{1+ \lbr \frac{\tau_S\tau_N\omega_D}{\tau_S+\tau_N} \rbr^2}
\right]^{\tau_N /\tau_S}.
\ee

The relation between the order parameter $\D$ and the minigap $E_g$ is given
by Eq.~(\ref{small}), which immediately yields
\be \label{E_g}
\frac{E_g}{\D_{BCS}} = \left[ \frac{\D_{BCS}}{2\omega_D}
\sqrt{1+ \lbr \frac{\tau_S\tau_N\omega_D}{\tau_S+\tau_N} \rbr^2}
\right]^{\tau_N /\tau_S}.
\ee
In the limit of a perfect interface (the Cooper limit), which is defined by
the condition $\tau_S\tau_N\omega_D/ (\tau_S+\tau_N) \ll 1$, Eq.~(\ref{E_g})
reproduces classical Cooper's result~\cite{Cooper} generalized to the
case of different Fermi parameters in the {\it S} and {\it N}
layers:~\cite{deGennes_review}
\be
E_g (\r\to 0)= 2\omega_D \exp \lbr -\frac 2{\left< \nu_0 \lambda \right>} \rbr,
\ee
with the effective interaction parameter
\be
\left< \nu_0 \lambda \right> =\frac{\tau_S}{\tau_S+\tau_N} \,\nu_{0S} \lambda.
\ee
This parameter can be considered as a result of averaging with the weighting
factors $\tau$, which are proportional to the {\it global} normal-metal DOS
per unit interval of energy, $\tau \propto {\cal A} d \nu_0$ (note that the
interaction parameter is zero in the {\it N} layer).

At the same time, we would like to emphasize that the Anderson limit does not
reduce to the Cooper limit with small corrections. On the contrary, due to the
relation $\D \ll \omega_D$, the Cooper limit's condition is {\it not}
satisfied over the most part of the Anderson limit's validity range;
therefore, the minigap $E_g$ and the quantities calculated below differ
drastically from the Cooper limit expressions.

Now we proceed to calculate the density of the superconducting electrons in
the {\it S} layer $n_S$. On this way, we immediately encounter the problem
that the above solution~(\ref{less_1}) of Eq.~(\ref{main}) at $E< \D$ is not
accurate enough for our purpose. In fact, as we will see below, the principal
contribution to the integral~(\ref{n}) determining $n_S$ comes from a narrow
region of energies near $E_g$. At the same time, Eq.~(\ref{less_1}) yields
$\Im\,[\sin^2 \theta_S] =0$ and hence no contribution at all from $E<\D$.
We thus have to calculate a correction to Eq.~(\ref{less_1}). Assuming this
correction to be small, we linearize Eq.~(\ref{main}) and obtain
\be
\sin\theta_S = \frac{iE_g}{\sqrt{ E^2 -E_g^2}} + X(E),
\ee
with
\be
X(E)= \frac{\tau_N E_g}{\lbr 1+\SN \rbr \left[ 1-\lbr \frac{E_g}E \rbr^2
\right]^2},
\ee
which is valid for all $E<\D$ except for a narrow vicinity of $E_g$.

The accurate consideration of the minigap's vicinity is possible due to the
fact that $|Z| \ll 1$ in this region. We define a new dimensionless quantity
$\epsilon$ as
\be
\frac E{E_g} = 1+ \frac{\lbr \tau_S \D \rbr^{2/3}}{2\lbr 1+\frac{\tau_S}{\tau_N}
\rbr^{4/3}} \, \epsilon ,
\ee
and consider the region $|E-E_g| \ll E_g$. The function
$\Im\,[\sin^2\theta_S]$, which determines $n_S$, has a peak at $\epsilon\sim
1$. An analysis of the coefficients~(\ref{coeff}) shows that only the terms
that contain zeroth, first, and third order in $Z$ should be retained in
Eq.~(\ref{main_Z}). Then, after rescaling
\be
Z=\frac{\lbr\tau_S\D\rbr^{1/3}}{\lbr 1+\frac{\tau_S}{\tau_N} \rbr^{2/3}} \, Y,
\ee
we obtain a cubic equation for the function $Y(\epsilon)$:
\be
4 Y^3 -\epsilon Y +i=0,
\ee
which can be solved analytically.

The density of the superconducting electrons $n_S$ at zero temperature is now
readily calculated:
\bea
\frac{n_S}{n_S^{BCS}} &=& \frac{E_g}{\D_{BCS}} \left\{ 1+\frac{11 \lbr\tau_S
\D \rbr^{5/6}}{2\pi \lbr 1+\SN \rbr^{4/3}} \right. \nonumber\\
&&\left. + \frac{2\tau_S \D}{\pi \lbr 1+\SN
\rbr^2} \left[ \NS +2\ln\frac{1+\SN}{\tau_S \D} \right] \right\},
\label{n_S_anderson}
\eea
with $E_g$ given by Eq.~(\ref{E_g}). The first term in the curly brackets is
the principal one; the two other terms become comparable to unity only near
the upper limit of applicability of Eq.~(\ref{n_S_anderson}).

The critical temperature of the bilayer $T_c$ is determined by
Eq.~(\ref{T_c}). In the Anderson limit, $(\tau_S +\tau_N)/ \tau_S \tau_N \gg
T_c$, and, with the use of the asymptotic form of the digamma function
$\psi(x) \sim \ln x$ at $x\gg 1$, we obtain
\be
\frac{T_c}{T_c^{BCS}} = \frac{E_g}{\D_{BCS}},
\ee
with $E_g$ given by Eq.~(\ref{E_g}). Interestingly, this result explains a
discrepancy in the formulas for $T_c$ of a thin bilayer that were found by
Cooper~\cite{Cooper} and McMillan.~\cite{McMillan} This discrepancy is
discussed in the classical paper by McMillan.~\cite{McMillan,Gamma} We
conclude that both cited results are correct, but their applicability
ranges are different, although within the Anderson limit. Cooper's result
corresponds to a perfect interface, $\tau_S\tau_N\omega_D/ (\tau_S+\tau_N)
\ll 1$, whereas McMillan's formula applies in the case $\tau_S\tau_N
\omega_D/ (\tau_S+\tau_N) \gg 1$.

Now we can discuss the general structure of the theory describing the bilayer
in the Anderson limit. In the limit $\r\to 0$, our results for the pairing
angle $\theta$ (which is constant over the entire bilayer, $\theta\equiv
\theta_S = \theta_N$) yield expressions which can be obtained from the BCS
ones [Eqs.~(\ref{sin_cos_BCS}), (\ref{Im_sin_sq_BCS})] if we substitute
the BCS order parameter $\D_{BCS}$ by the bilayer's minigap $E_g$. At $\r >0$,
corrections to this simple result are small while the Anderson limit's
conditions are satisfied. Therefore, we obtain a BCS-type theory with $E_g$
substituting $\D_{BCS}$ in all formulas.

The results of this Section immediately explain the numerical results in the
limit of relatively small $\r$, shown in Fig.~\ref{fig:narr_rho}. As we have
found, the Anderson limit implies the following relations between the
quantities under discussion:
\bea
E_g &=& \frac\pi\gamma T_c,\\
n_S &=& \pi \frac{m\sigma_S}{e^2} E_g,
\eea
which substitute Eqs.~(\ref{T_c_BCS}) and~(\ref{n_BCS}). For the Ta/Au
bilayer to which the numerical results refer, the Anderson limit is valid at
$\r <80$ (we see that the values of $\r$ can be large although they are {\it
relatively} small). Therefore, approaching $\r =80$, the curves $n_S /
n_S^{BCS}$ and $T_c / T_c^{BCS}$ tend to coincide, and $\D / \D_{BCS}$ exceeds
them by the large factor $(1+\tau_N / \tau_S) \approx 15$.

The temperature dependence of $\D$ and $n_S$, shown in
Fig.~\ref{fig:t_depend}, is quite different at $\r=90$ and $\r=110$; at
$\r=90$, the curves are much closer to the BCS behavior. This is also
explained by approaching the Anderson limit, where the curves coincide with
the BCS ones.

The DOS in the {\it S} and {\it N} layers coincide in the Anderson limit
[Eqs.~(\ref{cos}), (\ref{less_1})], showing the standard BCS-like peculiarity
at $E=E_g$. The tendency to such behavior is illustrated by the DOS curve for
$\r=80$ in Fig.~\ref{fig:dos}.

% 666666666666666666666666666666666666666666666666666666666666666666666666666
\section{Parallel critical field} \label{sec:H_c}

We proceed to calculate the critical magnetic field $H_c$ directed along the
plane of the bilayer. As it was mentioned in Sec.~\ref{subsec:theta}, in the
presence of an external magnetic field, the superconducting phase gradient in
the Usadel equations~(\ref{usadel_theta}) must be substituted by its gauge
invariant form, which can be expressed via the supercurrent velocity
${\bf v}$. The spatial distribution of ${\bf v}$ in the bilayer can be found
as follows.

Let us direct the $x$-axis along the magnetic field ${\bf H}$.
The supercurrents ${\bf j}=-en{\bf v}$ are directed along the bilayer and
perpendicularly to ${\bf H}$, {\it i.e.}, ${\bf j} =(0,j(z),0)$ and ${\bf v} =
(0,v(z),0)$. Near $H_c$, the magnetic field inside the bilayer is uniform, so
the vector potential can be chosen as ${\bf A}=(0,-zH,0)$. The
supercurrent velocity distribution is determined by the equation $\nabla\times
{\bf v}= e{\bf H}/m$. Another essential point is the continuity of ${\bf v}$
at the {\it SN} interface, which follows from the continuity of the
superconducting phase $\varphi$ [see the boundary condition~(\ref{b_3})].
The result is
\be \label{v}
v(z) = v_0-\frac{eH}m z,
\ee
where $v_0$ is the supercurrent velocity at the interface, which must be
determined from the condition that the total charge transfer across the
bilayer's cross-section is zero:
\be
\int_{-d_N}^{d_S} j(z)\, dz=0,
\ee
leading to
\be \label{v_0}
v_0 =\lbr\frac{eH}{2m}\rbr \frac{n_S d_S^2 -n_N d_N^2}{n_S d_S + n_N d_N}.
\ee
The density of the superconducting electrons is constant in each layer ($n_S$
and $n_N$).

Near $H_c$, the superconducting correlations are small, $|\theta| \ll 1$, and
the Usadel equation~(\ref{usadel_theta_1}) for the paring angle $\theta(z,E)$
can be linearized:
\bml \label{usadel_H_c}
\bea
\frac{D_N}2 \frac{\partial^2\theta_N}{\partial z^2}
+\lbr iE -2m^2 D_N\, {\bf v}^2 \rbr\theta_N &=& 0,\\
\frac{D_S}2 \frac{\partial^2\theta_S}{\partial z^2}
+\lbr iE -2m^2 D_S\, {\bf v}^2 \rbr\theta_S+ \left| \D \right| &=& 0.
\eea
\eml
At the same time, the second Usadel equation~(\ref{usadel_theta_2}) is
trivial: its l.h.s. is proportional to
\be
\nabla \lbr \sin^2 \theta\,\, {\bf v} \rbr =\sin 2\theta\,\, \nabla\theta\,\,
{\bf v} + \sin^2 \theta\,\, \nabla {\bf v},
\ee
where both terms vanish due to the fact that $\nabla\theta$ is directed
along the $z$-axis whereas ${\bf v}$ is parallel to the $y$-axis.

The pairing angle $\theta$ is almost spatially constant in each layer; this
allows us to average each of Eqs.~(\ref{usadel_H_c}) over the thickness of
the corresponding layer, obtaining
\bml \label{after_avr}
\bea
\left. \frac{\partial\theta_N}{\partial z} \right|_{z=0} &=&
\frac{2d_N}{D_N} \lbr E_N -iE \rbr \theta_N,\\
\left. \frac{\partial\theta_S}{\partial z} \right|_{z=0} &=&
\frac{2d_S}{D_S} \left[ \lbr iE -E_S \rbr\theta_S+ \left| \D \right| \right],
\eea
\eml
where
\bea
E_N &=& 2m^2 D_N \left< {\bf v}^2 (z) \right>_N, \nonumber \\
E_S &=& 2m^2 D_S \left< {\bf v}^2 (z) \right>_S
\eea
are $H$-dependent energies. Using Eqs.~(\ref{v}), (\ref{v_0}), we express them
via $H_c$ and the densities of the superconducting electrons:
\be
\label{E_S}
E_S = \frac{D_S\, e^2 H_c^2}6 \left[ d_S^2+ 3 d_N^2 \frac{n_N^2 \lbr d_S+d_N \rbr^2}
{\lbr n_S d_S + n_N d_N \rbr^2} \right],
\ee
and $E_N$ is obtained by the interchange of all the {\it S} and {\it N}
indices.

Substituting~(\ref{after_avr}) into the boundary
conditions~(\ref{gran_uslovija}) (which should be linearized), we find
\bml \label{theta_for_H_c}
\bea
\theta_N &=&\tau_S \left| \D \right| \left/ \left\{ \tau_S E_S+\tau_N E_N +
\tau_S \tau_N E_S E_N -\tau_S \tau_N E^2 \right. \right. \nonumber\\
&&\phantom{\tau_S \left| \D \right| / \bigl\{ } \left. -iE\left[
\tau_S+\tau_N + \tau_S\tau_N \lbr E_S+E_N \rbr \right] \right\},\\
\theta_S &=& \lbr 1+\tau_N E_N -i\tau_N E \rbr \theta_N.
\eea
\eml
The order parameter $\D$ cancels out from the self-consistency
equation~(\ref{Delta}). However, the resulting equation alone does not suffice
for determining $H_c(T)$ because it contains $E_S$ and $E_N$, which are
functions of $n_N/n_S$. Therefore, to obtain a closed system, we must consider
the self-consistency equation together with the equation determining the
ratio $n_N/n_S$; the latter
equation is obtained from Eq.~(\ref{n}). The resulting system of two nonlinear
equations for the quantities $H_c$ and $n_N/n_S$ is
\bml \label{system}
\bea \label{system_1}
\ln\frac{2\omega_D}{\D_{BCS}} &=& \int_0^{\omega_D} dE\, \tanh\!\lbr\!
\frac E{2T} \!\rbr \frac{\Im\theta_S}{\left| \D \right|},\\
\label{system_2}
\frac{n_N}{n_S} &=& \frac{\sigma_N \int_0^\infty dE\, \tanh\!\lbr \frac E{2T}
\rbr \Im\theta_N^2}{\sigma_S \int_0^\infty dE\, \tanh\!\lbr \frac E{2T} \rbr
\Im\theta_S^2},
\eea
\eml
with $\theta_N$ and $\theta_S$ given by
Eqs.~(\ref{theta_for_H_c}). The first equation of the system,
Eq.~(\ref{system_1}), can be written via the digamma functions, thus taking
exactly the same form as Eq.~(\ref{digamma}) below (which determines the
perpendicular upper critical field) if we denote ${\cal E}_S = E_S +1/\tau_S$,
${\cal E}_N = E_N +1/\tau_N$.

In the limit $\r\to\infty$, Eqs.~(\ref{system}) lead to the BCS result. In
this case, the layers uncouple, the density of the superconducting electrons
in the {\it N} layer vanishes, $n_N /n_S \to 0$, and Eq.~(\ref{system_1})
finally yields
\be \label{like_MdG}
\ln\frac{T_c^{BCS}}T = \psi \lbr \frac 12 +\frac{D_S \left[ eH_c^{BCS} d_S
\right]^2}{12\pi T} \rbr -\psi \lbr \frac 12 \rbr,
\ee
which determines the parallel critical field $H_c^{BCS} (T)$ of a thin
superconducting film. Another immediate consequence of Eqs.~(\ref{system}) is
the critical temperature of the bilayer $T_c$, which can be found from the
condition $H_c (T_c) =0$: in this case, Eqs.~(\ref{system}) reproduce
Eq.~(\ref{T_c}).
% -----------------------------   FIGURE   ----------------------------------
\begin{figure}
%\centerline{\epsfxsize=8.7cm \epsfbox{h_c.eps}}
\centerline{\epsfxsize=\hsize \epsfbox{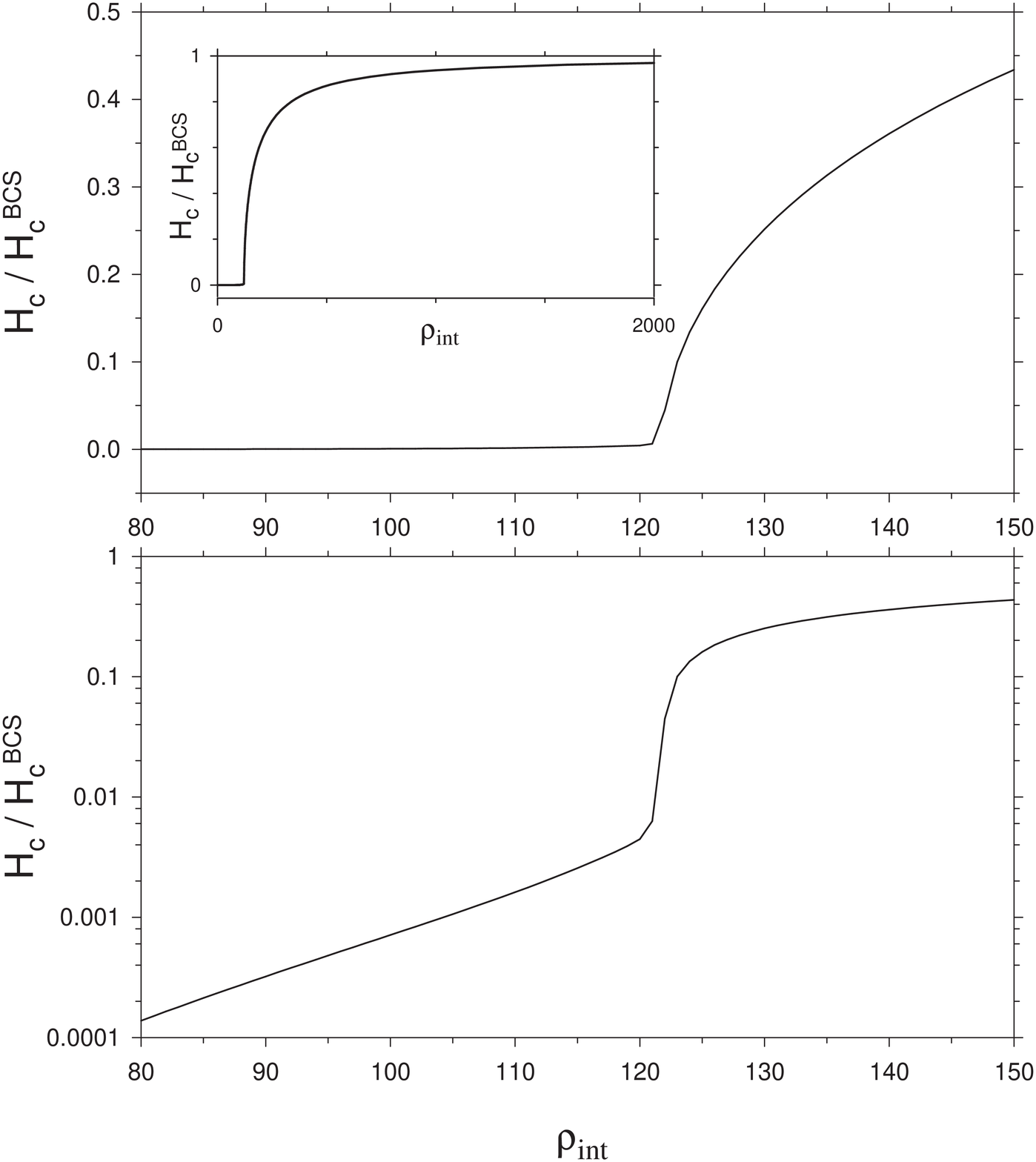}} \vspace{2mm}
\caption{Parallel critical field $H_c$, normalized by the BCS value, versus
$\r$ at zero temperature. The upper and lower graphs differ only in the
scaling of the ordinate axis (normal and logarithmic, respectively). The
nature of the steep behavior of $H_c$ at $\r =120$--$123$, which is best seen
from the lower graph, is explained in the text. The inset shows $H_c (\r)$ on
a wider scale over $\r$.}
\label{fig:h_c}
\end{figure}
% -----------------------------   FIGURE   ----------------------------------
\begin{figure}
%\centerline{\epsfxsize=8.6cm \epsfbox{h_c_vs_t.eps}}
\centerline{\epsfxsize=\hsize \epsfbox{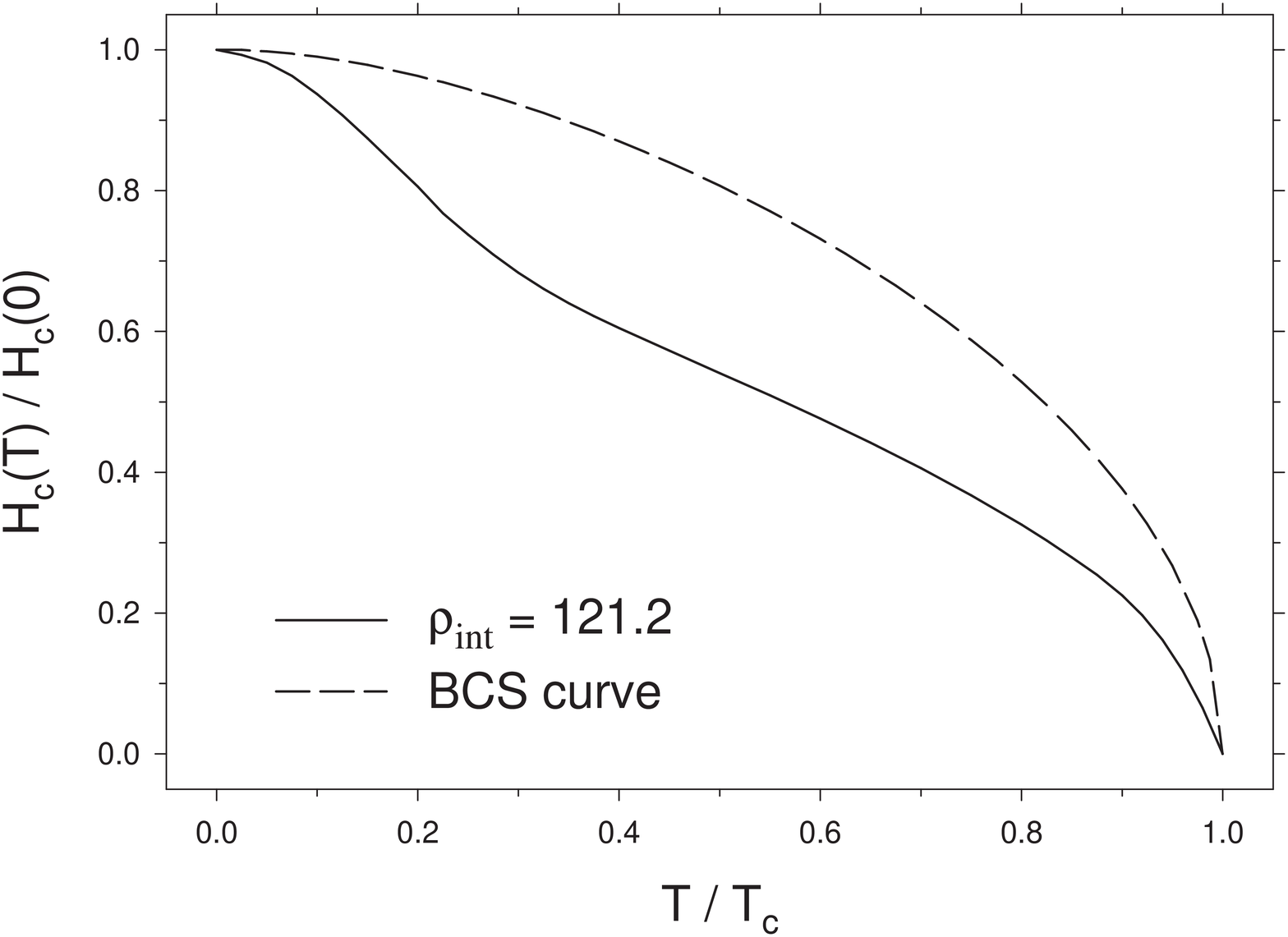}} \vspace{2mm}
\caption{Temperature dependence of the parallel critical field $H_c$ at $\r=
121.2$. The experimental value of $H_c(0)$, analyzed with the use of the
results shown in Fig.~\ref{fig:h_c}, suggests that this value of $\r$
corresponds to the experiment by Kasumov {\it et al.}~\protect\cite{Kasumov}
The critical field is normalized by its zero-temperature value, and the
temperature is normalized by the corresponding $T_c$. For comparison, the same
dependence is plotted for the BCS case.}
\label{fig:h_c_vs_t}
\end{figure}
%\vspace{-2mm}
% -----------------------------   FIGURE   ----------------------------------
\begin{figure}
%\centerline{\epsfxsize=8.6cm \epsfbox{z_0.eps}}
\centerline{\epsfxsize=\hsize \epsfbox{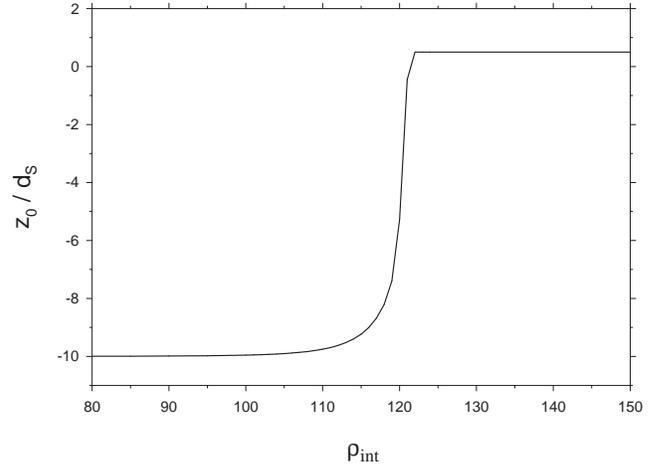}} \vspace{1mm}
\caption{Position of the stationary point $z_0$ of the supercurrent
distribution versus $\r$ at zero temperature. The coordinate $z_0$ is
normalized by the {\it S} layer thickness $d_S$. The fast shift in $z_0$ from
the center of the {\it S} layer at large $\r$ to (nearly) the center of the
{\it N} layer at small $\r$ corresponds to the steep drop in $H_c$, shown in
Fig.~\ref{fig:h_c}.}
\label{fig:z_0}
\end{figure}

The system of equations~(\ref{system}) can be solved numerically at arbitrary
values of the temperature $T$ and the interface resistance $\r$; the results
for $H_c$ are presented in Figs.~\ref{fig:h_c}, \ref{fig:h_c_vs_t}.

A remarkable feature of the function $H_c (\r)$ at zero temperature
(Fig.~\ref{fig:h_c}) is the steep behavior of $H_c$ at $\r= 120$--$123$. This
feature is due to rearrangement of the supercurrents inside the bilayer, which
occurs in the following way. The supercurrent velocity changes across the
thickness of the bilayer according to the simple linear law~(\ref{v}). This
supercurrent distribution may be characterized by the position of the
stationary point $z_0$, where the supercurrent velocity is zero: $v(z_0) =0$,
hence $z_0 = m v_0 /eH$. At large values of the interface resistance $\r$, the
density of the superconducting electrons in the {\it N} layer is very small,
$n_N /n_S \ll 1$, and the supercurrents circulate only in the {\it S} part of
the system; this case corresponds to
\be
z_0 = \frac{d_S}2.
\ee
Then, while decreasing $\r$, a shift in $z_0$ occurs. Now the supercurrents
in the {\it S} layer are not compensated (in the sense of the charge
transfer); therefore, they must be compensated by the supercurrents in the
{\it N} layer, which are enhanced due to significant increase in $n_N$. This
situation corresponds to the beginning of the drop in $H_c$. The ratio of the
superconducting electrons' densities grows rapidly, approaching the Anderson
limit value $n_N/n_S = \sigma_N/\sigma_S$ (see Sec.~\ref{subsec:H_c_Anderson}
below); simultaneously, $z_0$ tends to
\be \label{z_0}
z_0 = \frac{\sigma_S d_S^2 -\sigma_N d_N^2}{2 \lbr \sigma_S d_S + \sigma_N
d_N \rbr},
\ee
and the steep drop in $H_c$ finishes. For the bilayer to which the numerical
results refer, $d_S \ll d_N$ and $\sigma_S \ll \sigma_N$, so Eq.~(\ref{z_0})
yields $z_0 \approx -d_N /2$.

This scenario is illustrated by Fig.~\ref{fig:z_0}, which has been obtained
numerically.

The analytical solution of Eqs.~(\ref{system}) at zero temperature in the
Anderson limit is presented below.

\subsection{$H_c$ at zero temperature in the Anderson limit}
\label{subsec:H_c_Anderson}

In the zero-temperature Anderson limit (defined by the conditions $\tau_S
E_S$, $\tau_N E_N \ll 1$), the ratio of the superconducting electrons'
densities~(\ref{system_2}) becomes independent of the magnetic field, $n_N/n_S
=\sigma_N/\sigma_S$, and the self-consistency equation~(\ref{system_1}) yields
\be \label{H_c}
\frac{\tau_S E_S +\tau_N E_N}{\tau_S +\tau_N} = \frac{\D_{BCS}}2 \left[
\frac{\D_{BCS}}{2\omega_D} \sqrt{1+ \lbr \frac{\tau_S\tau_N\omega_D}
{\tau_S+\tau_N} \rbr^2} \right]^\NS
\ee
which determines $H_c$. This result can be compared to the BCS case, which
corresponds to the limit $\r\to\infty$. In this case, the density of the
superconducting electrons in the {\it N} layer vanishes, $n_N /n_S \to 0$, and
the self-consistency equation yields
\be \label{H_c_BCS}
E_S^{BCS} = \frac{\D_{BCS}}2,
\ee
where $E_S^{BCS}$ is given by Eq.~(\ref{E_S}) with $n_N=0$. Finally,
\be \label{Hc_BCS}
H_c^{BCS} = \frac{\sqrt 3 \Phi_0}{\pi\, \xi_{BCS}\, d_S},\qquad \xi_{BCS} =
\sqrt{\frac{D_S}{\D_{BCS}}},
\ee
where $\Phi_0 = \pi/e$ is the flux quantum, and $\xi_{BCS}$ is the correlation
length in the dirty limit.

Remarking that the r.h.s. of Eq.~(\ref{H_c}) is identical to $E_g/2$ with the
minigap $E_g$ given by Eq.~(\ref{E_g}), we see that equation~(\ref{H_c}),
determining the parallel critical field of the bilayer in the Anderson limit,
is obtained from
the BCS equation~(\ref{H_c_BCS}) if we substitute the order parameter
$\D_{BCS}$ by the minigap $E_g$ (in accordance with the results of
Sec.~\ref{sec:anderson}) and the $H$-dependent energy $E_S^{BCS}$ by the
corresponding averaged quantity $(\tau_S E_S +\tau_N E_N) / (\tau_S +\tau_N)$.

The explicit result for the parallel critical field of the bilayer,
obtained from Eq.~(\ref{H_c}), can be cast into a BCS-like form:
\be \label{H_c_BCS_like}
H_c = \frac{\sqrt 3 \Phi_0}{\pi \xi d_{\rm eff}}.
\ee
The bilayer's correlation length $\xi$ is the characteristic space scale on
which the order parameter (or the pairing angle $\theta$, or the Green
function) varies in the absence of the magnetic field. In the Anderson limit
(under discussion), the explicit formula
for $\xi$ is a natural generalization of the BCS expression [see
Eq.~(\ref{Hc_BCS})] which implies that $D_S$ must be substituted by the
averaged diffusion constant $\left< D \right>$ and $\D_{BCS}$ must be
substituted (in accordance with the results of Sec.~\ref{sec:anderson}) by the
bilayer's characteristic energy scale, the minigap $E_g$ [Eq.~(\ref{E_g})]:
\be \label{xi}
\xi=\sqrt{\frac{\left< D \right>}{E_g}},\qquad \left< D \right> =
\frac{\tau_S D_S+\tau_N D_N}{\tau_S +\tau_N}.
\ee
The effective thickness of the bilayer in Eq.~(\ref{H_c_BCS_like}) is
\bea
&&d_{\rm eff} = \Bigl[ \lbr \sigma_S d_S +\sigma_N d_N \rbr \lbr \sigma_S d_S^3
+ \sigma_N d_N^3 \rbr \nonumber \\
&&\left. + 3\sigma_S \sigma_N d_S d_N \lbr d_S+d_N \rbr^2 \Bigr]^{1/2}
\right/ \lbr \sigma_S d_S + \sigma_N d_N \rbr .
\eea
In the case of equal conductivities, $\sigma_S = \sigma_N$, the effective
thickness is simply the geometrical one: $d_{\rm eff} = d_S + d_N$. This case
corresponds to a uniform density of the superconducting electrons, $n_S =
n_N$, which implies a continuous distribution of the supercurrents, centered
at the middle of the bilayer [this can be also seen from Eq.~(\ref{z_0}) which
yields $z_0 =(d_S+d_N)/2$ in the case $\sigma_S=\sigma_N$]. However, in a more
subtle situation when the conductivities are different, the density of the
supercurrent experiences a jump at the {\it SN} interface; this nontrivial
supercurrent distribution results in the nonequivalence of $d_{\rm eff}$ to
the geometrical thickness of the bilayer.

% 777777777777777777777777777777777777777777777777777777777777777777777777777
\section{Perpendicular upper critical field} \label{sec:H_c2}

Now we turn to calculating the upper critical field $H_{c2}$ perpendicular to
the plane of the bilayer.

As in the case of the parallel critical field, we start with discussing the
supercurrent distribution, which is now a function of the sample boundaries in
the $xy$-plane, perpendicular to the magnetic field ${\bf H}$ (the magnetic
field is directed along the $z$-axis). The infinite bilayer under
consideration can
be thought of as a disk of a large radius; let us assume $x=0$, $y=0$ at the
axis of the disk. Then the supercurrent distribution is axially symmetric,
and, with the gauge chosen as ${\bf A}=\left[ {\bf Hr} \right]/2$, the
superconducting phase must be constant, $\varphi=0$, which yields a simple
result for the supercurrent velocity: ${\bf v}=e{\bf A}/m$.

Near $H_{c2}$, the superconducting correlations are small, $|\theta| \ll 1$,
and the Usadel equations can be linearized:
\bml
\bea
\label{usadel_h_c2}
-\frac D2 \lbr -i\nabla +2e{\bf A} \rbr^2 \theta +iE\theta +\D &=& 0, \\
{\bf A} \nabla\theta &=& 0.
\eea
\eml
The second of these equations is trivially satisfied because $\theta({\bf r})$
is axially symmetric.

Thus, the Usadel equations reduce to the single Eq.~(\ref{usadel_h_c2}) for
the pairing angle $\theta({\bf r},E)$. Introducing the cylindrical coordinates
${\bf r} \leftrightarrow (z,\bbox\rho)$ and denoting $\hat {\bf P} =-i
\nabla_{\bbox\rho} +2e{\bf A}(\bbox\rho)$, we rewrite this equation as
\bml \label{usadel_detailed}
\bea
\frac{D_N}2 \frac{\partial^2 \theta_N}{\partial z^2} -\frac{D_N}2 \hat
{\bf P}^2 \theta_N + iE\theta_N &=& 0,\\
\frac{D_S}2 \frac{\partial^2 \theta_S}{\partial z^2} -\frac{D_S}2 \hat
{\bf P}^2 \theta_S + iE\theta_S +\D &=& 0.
\eea
\eml
We cannot solve these equations straightforwardly because near the upper
critical field, the order parameter $\D({\bbox \rho})$ is a nontrivial unknown
function of the in-plane coordinate $\bbox\rho$ (while the $z$-dependence is
absent due to the small thickness of the bilayer). In this situation, we
employ the following approach.

Averaging each of Eqs.~(\ref{usadel_detailed}) over the thickness of the
corresponding layer, we obtain
\bml \label{after_avrg}
\bea
\left. \frac{\partial\theta_N}{\partial z} \right|_{z=0} &=&
\frac{2d_N}{D_N} \lbr \frac{D_N}2 \hat {\bf P}^2 \theta_N -iE\theta_N \rbr,\\
\left. \frac{\partial\theta_S}{\partial z} \right|_{z=0} &=&
\frac{2d_S}{D_S} \lbr -\frac{D_S}2 \hat {\bf P}^2 \theta_S +iE\theta_S +\D
\rbr .
\eea
\eml
The averaged pairing angles entering the r.h.s. of Eqs.~(\ref{after_avrg})
are
\bml
\bea
\theta_N (\bbox\rho,E) &=& \frac 1{d_N} \int_{-d_N}^0 dz\, \theta_N
(z,\bbox\rho,E),\\
\theta_S (\bbox\rho,E) &=& \frac 1{d_S} \int_0^{d_S} dz\, \theta_S
(z,\bbox\rho,E).
\eea
\eml
Substituting Eqs.~(\ref{after_avrg}) into the boundary
conditions~(\ref{gran_uslovija}) (which should be linearized), we obtain a
system of two differential equations for the function $\theta(\bbox\rho,E)$:
\bea
\tau_N \lbr \frac{D_N}2 \hat {\bf P}^2 \theta_N -iE\theta_N \rbr\!
&=& \tau_S \lbr -\frac{D_S}2 \hat {\bf P}^2 \theta_S +iE\theta_S
+\D \rbr \nonumber\\ &=& \theta_S-\theta_N.
\label{gr_us}
\eea

From the vicinity of the superconductive transition it follows that the
pairing angle $\theta$ depends on the order parameter $\D$ linearly:
\bml
\bea
\theta_N (\bbox\rho,E) &=& \frac{\D \lbr\bbox\rho\rbr}{\alpha_N\lbr E\rbr},\\
\theta_S (\bbox\rho,E) &=& \frac{\D \lbr\bbox\rho\rbr}{\alpha_S \lbr E\rbr},
\eea
\eml
where the functions $\alpha_N (E)$ and $\alpha_S (E)$ are spatially
independent. Then Eqs.~(\ref{gr_us}) can be rewritten as
\bml
\be
\frac{D_N}2 \hat {\bf P}^2 \D (\bbox\rho) = \left[ iE + \frac1{\tau_N}
\lbr \frac{\alpha_N \lbr E \rbr}{\alpha_S \lbr E \rbr} -1 \rbr \right]
\D (\bbox\rho),
\ee
\be
\frac{D_S}2 \hat {\bf P}^2 \D (\bbox\rho) = \left[ iE + \frac 1{\tau_S}
\lbr \frac{\alpha_S \lbr E \rbr}{\alpha_N \lbr E \rbr} -1 \rbr + \alpha_S (E)
\right] \D (\bbox\rho).
\ee
\eml

We see that the order parameter must be an eigenfunction of the differential
operator
$\hat {\bf P}^2$. Moreover, in order to obtain the largest value of $H_{c2}$,
we should choose the eigenfunction corresponding to the lowest eigenvalue (in
complete analogy with Refs.~[\onlinecite{Abrikosov,HW}]). The solution of the
emerging eigenvalue problem is readily found thanks to its formal equivalence
to the problem of determining the Landau levels of a two-dimensional particle
with the ``mass'' $1/D$ and the charge $-2e$ in the uniform magnetic field
${\bf H}$ directed along the third dimension. The lowest Landau level is
$DeH$; the function $\alpha_S (E)$ is straightforwardly determined,
\be
\alpha_S (E) = D_S\, eH -iE +\frac{\tau_N \lbr D_N\, eH -iE \rbr}{\tau_S
\left[ 1+\tau_N \lbr D_N\, eH -iE \rbr \right]},
\ee
and we substitute $\theta_S (\bbox\rho,E)$ into the self-consistency
equation~(\ref{Delta}). The order parameter $\D (\bbox\rho)$ cancels out, and
the resulting equation, which determines $H_{c2}(T)$, can be cast into the
form
\end{multicols}
\widetext
\noindent
\rule{87mm}{0.4pt}\rule{0.4pt}{3mm}
\bea
\ln\frac{T_c^{BCS}}T &=&  -\frac{\tau_N}{\tau_S+\tau_N}
\ln\sqrt{1+\lbr \frac{\tau_S+\tau_N}{\tau_S\tau_N\omega_D} \rbr^2}
-\psi \lbr \frac 12 \rbr \nonumber \\
&& +\frac 12 \left[ 1+\frac{{\cal E}_S-{\cal E}_N}{\sqrt{\lbr {\cal E}_S-{\cal E}_N
\rbr^2 + 4/\tau_S\tau_N}} \right] \psi \lbr \frac 12 +\frac 1{4\pi T}
\left[ {\cal E}_S+{\cal E}_N +\sqrt{\lbr {\cal E}_S-{\cal E}_N \rbr^2 +
\frac 4{\tau_S\tau_N}} \right] \rbr \nonumber \\
&& +\frac 12 \left[ 1-\frac{{\cal E}_S-{\cal E}_N}{\sqrt{\lbr {\cal E}_S-{\cal E}_N
\rbr^2 + 4/\tau_S\tau_N}} \right] \psi \lbr \frac 12 +\frac 1{4\pi T}
\left[ {\cal E}_S+{\cal E}_N -\sqrt{\lbr {\cal E}_S-{\cal E}_N \rbr^2 +
\frac 4{\tau_S\tau_N}} \right] \rbr ,
\label{digamma}
\eea
\hfill\rule[-3mm]{0.4pt}{3mm}\rule{87mm}{0.4pt}
\begin{multicols}{2}
\narrowtext
% -----------------------------   FIGURE   ----------------------------------
\begin{figure}
%\centerline{\epsfxsize=8.7cm \epsfbox{h_c2.eps}}
\centerline{\epsfxsize=\hsize \epsfbox{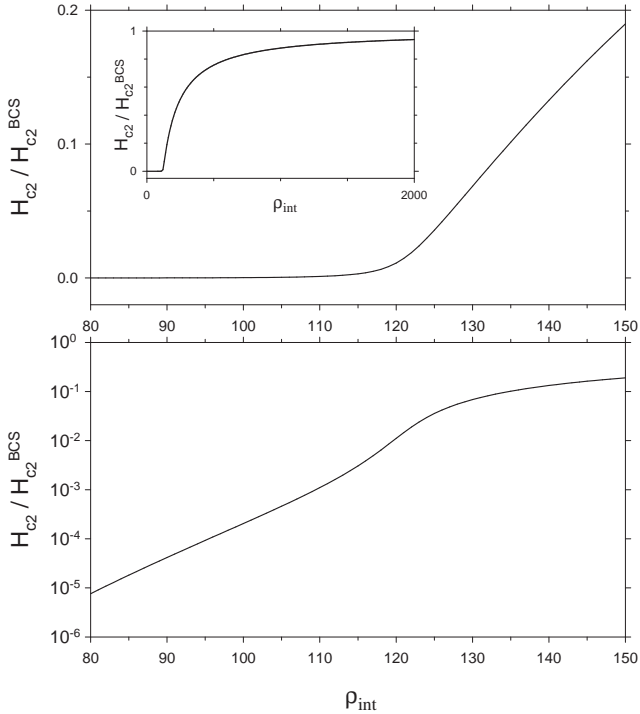}} \vspace{2mm}
\caption{Perpendicular upper critical field $H_{c2}$, normalized by its BCS
value, versus $\r$ at zero temperature. The upper and lower graphs differ only
in the scaling of the ordinate axis (normal and logarithmic, respectively).
The inset shows $H_{c2} (\r)$ on a wider scale over $\r$.}
\label{fig:h_c2}
\end{figure}
\vspace{2mm}
% -----------------------------   FIGURE   ----------------------------------
\begin{figure}
%\centerline{\epsfxsize=9cm \epsfbox{h_c2_vst.eps}} \vspace{2mm}
\centerline{\epsfxsize=\hsize \epsfbox{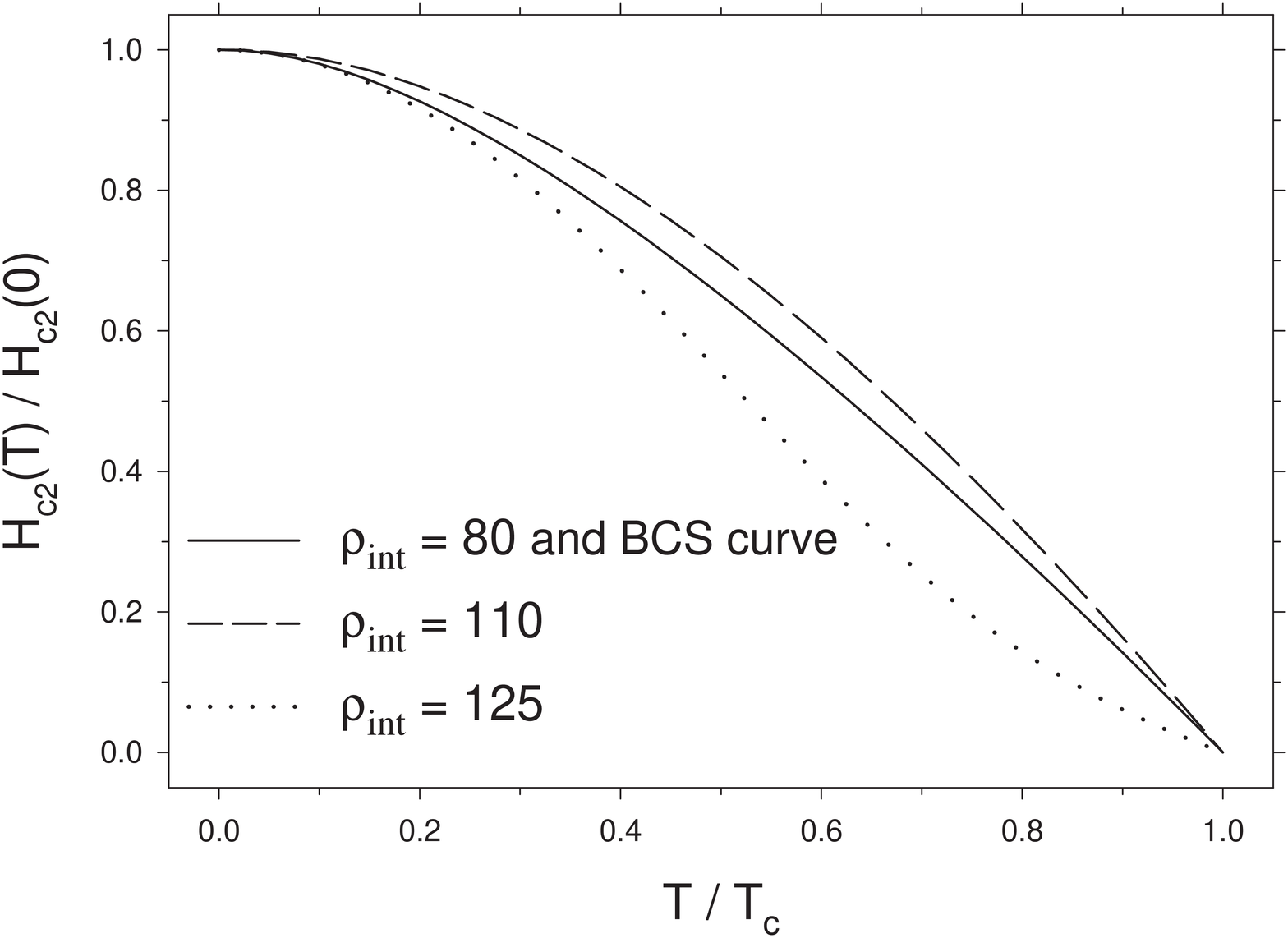}} \vspace{2mm}
\caption{Temperature dependence of the perpendicular upper critical field
$H_{c2}$ at $\r=80$, 110, 125, and in the BCS case. In each case, the critical
field is normalized by its zero-temperature value, and the temperature is
normalized by the corresponding $T_c$. According to the results of
Sec.~\ref{sec:anderson}, the curves in the BCS and Anderson ($\r=80$) limits
coincide. At intermediate values of $\r$, the curves can lie both above
($\r=110$) and below ($\r=125$) the BCS curve.}
\label{fig:h_c2_vst}
\end{figure}
\noindent
where
\bea
{\cal E}_S &=& D_S\, eH_{c2}+ \frac 1{\tau_S} , \nonumber \\
{\cal E}_N &=& D_N\, eH_{c2}+ \frac 1{\tau_N}
\eea
are $H$-dependent energies. The logarithmic term in the r.h.s. of
Eq.~(\ref{digamma}) takes account of the finiteness of the Debye energy
$\omega_D$; it becomes important only in the limit of a perfect interface (the
Cooper limit), {\it i.e.}, when $\tau_S\tau_N\omega_D /(\tau_S+\tau_N) \ll 1$.

In the limit $\r\to\infty$, Eq.~(\ref{digamma}) yields the classical result of
Maki~\cite{Maki} and de~Gennes~\cite{deGennes} for the BCS case,
\be \label{MdG}
\ln\frac{T_c^{BCS}}T = \psi \lbr \frac 12 +\frac{D_S\, eH_{c2}^{BCS}}{2\pi T}
\rbr -\psi \lbr \frac 12 \rbr,
\ee
which is valid for bulk superconductors and superconductive layers of
arbitrary thickness (when the magnetic field is directed perpendicularly to
them). Another immediate consequence of Eq.~(\ref{digamma}) is the critical
temperature of the bilayer $T_c$, which can be found from the condition
$H_{c2} (T_c) =0$: in this case, Eq.~(\ref{digamma}) reproduces
Eq.~(\ref{T_c}).

Equation~(\ref{digamma}) can be solved numerically at arbitrary values of the
temperature $T$ and the interface resistance $\r$; the results for $H_{c2}$
are presented in Figs.~\ref{fig:h_c2}, \ref{fig:h_c2_vst}.

The analytical solution of Eq.~(\ref{digamma}) at zero temperature in the
Anderson limit is presented below.

\subsection{$H_{c2}$ at zero temperature in the Anderson limit}

In the zero-temperature Anderson limit (defined by the conditions
$D_S\, eH_{c2} \ll 1/\tau_S$, $D_N\, eH_{c2} \ll 1/\tau_N$),
Eq.~(\ref{digamma}) yields
\be \label{H_c2}
\frac{H_{c2}}{H_{c2}^{BCS}} = \frac{\lbr \tau_S+\tau_N \rbr D_S}{\tau_S D_S +
\tau_N D_N} \left[ \frac{\D_{BCS}}{2\omega_D}
\sqrt{1+ \lbr \frac{\tau_S\tau_N\omega_D}{\tau_S+\tau_N} \rbr^2}
\right]^\NS
\ee
where the zero-temperature BCS value of the upper critical field, as follows
from Eq.~(\ref{MdG}), is
\be
H_{c2}^{BCS} =\frac{\D_{BCS}}{2eD_S} = \frac{\Phi_0}{2\pi\xi_{BCS}^2} .
\ee

It is instructive to rewrite the perpendicular upper critical field of the
bilayer~(\ref{H_c2}) in the standard BCS-like form
\be
H_{c2} =\frac{\Phi_0}{2\pi\xi^2},
\ee
where $\xi$ is the bilayer's correlation length given by Eq.~(\ref{xi})
[the physical interpretation of this result for $\xi$ precedes
Eq.~(\ref{xi})].

% 8888888888888888888888888888888888888888888888888888888888888888888
\section{{\it SNS}, {\it NSN}, {\it SNINS}, {\it NSISN}, and superlattices}
\label{sec:SNS}

Our results for $\D$, $n_S$, $T_c$, $E_g$, $\nu (E)$, and $H_{c2}$
({\it i.e.}, all the results except $H_c$) can be directly applied to more
complicated structures such as {\it SNS} and {\it NSN} trilayers, {\it SNINS}
and {\it NSISN} systems, and {\it SN} superlattices.

Let us consider, for example, a symmetric {\it SNS} trilayer consisting of two
identical {\it S} layers of thickness $d_S$ separated by a {\it N} layer of
thickness $2 d_N$. The {\it SN} interfaces can have arbitrary (but equal)
resistances. As before, the $z$-axis is perpendicular to the plane of the
structure. This trilayer can be imagined as composed of two identical bilayers
perfectly joined together along the {\it N} sides. Indeed, the pairing angle
$\theta$ has zero $z$-derivative on the outer surfaces of the bilayers, thus
producing the correct (symmetric in the $z$-direction) solution for $\theta$
in the resulting trilayer. Consequently, the symmetric {\it SNS} trilayer has
exactly the same physical properties [$\D$, $n_S$, $T_c$, $E_g$, $\nu (E)$,
$H_{c2}$] as the {\it SN} bilayer considered in the present paper. The only
point where the above reasoning fails is the calculation of the parallel
critical field $H_c$. In this case, the combination of the supercurrent
distributions in the two bilayers does not yield the correct distribution in
the resulting {\it SNS} trilayer, which implies that the Usadel equations for
the two systems are different.

Evidently, the above reasoning, based on the formal equivalence of the
outer-surface boundary condition for the bilayer to the symmetry-caused
condition in the middle of the {\it SNS} trilayer, also holds for symmetric
{\it NSN} trilayers ({\it N} layers of thickness $d_N$, {\it S} layer of
thickness $2 d_S$, identical {\it SN} interfaces) and {\it SN} superlattices
({\it N} layers of thickness $2 d_N$, {\it S} layers of thickness $2 d_S$,
identical {\it SN} interfaces). Moreover, the same applies to systems composed
of two bilayers in {\it nonideal} contact with each other: {\it SNINS} and
{\it NSISN} (where {\it I} stands for an arbitrary potential barrier), because
the presence of a potential barrier does not violate the applicability of the
symmetry argument. Thus, all the results obtained for the bilayer (except
$H_c$) are also valid for these structures.

% 9999999999999999999999999999999999999999999999999999999999999999999999999999
\section{Discussion} \label{sec:discussion}

An essential property of the bilayer used throughout the paper is its small
thickness. Now we shall argue that the bilayer studied in the experiment by
Kasumov {\it et al.}~\cite{Kasumov} (and to which our numerical results refer)
can be considered thin. The Usadel equations~(\ref{usadel_bi}) imply that
the characteristic space scale of the bilayer's properties variation is
$\sqrt{D_{N,S}/E_0}$ for the {\it N} and {\it S} layers, respectively.
However, the correct determination of the characteristic energy scale $E_0$ is
a nontrivial problem. Our results suggest that $E_0$ is always smaller than
the order parameter $\D$: in the BCS limit ($\r\to\infty$), $E_0$ approaches
$\D$, whereas in the opposite (Anderson) limit, $E_0$ is determined by the
minigap $E_g$ [see Eq.~(\ref{small})]. For the following discussion, it is
convenient to write the condition of the small thickness of the bilayer as
\be \label{thin}
d_{N,S} \ll \sqrt{\frac \D{E_0}} \, \sqrt{\frac{\D_{BCS}}\D} \,
\sqrt{\frac{D_{N,S}}{\D_{BCS}}}.
\ee
The individual layers' thicknesses are $d_N =$ 100~nm and $d_S =$ 5~nm. The
third multiplier (the BCS correlation length) in the r.h.s. of the
condition~(\ref{thin}) equals 194~nm and 16~nm for the {\it N} and {\it S}
layers, respectively. At the same time, each of the first two multipliers in
the r.h.s. of the condition~(\ref{thin}) exceeds unity. We can thus conclude
that the bilayer can indeed be considered thin.

Now we turn to a possible experimental application of our results. Our
results provide a method for determining $\r$, a very important parameter of
the bilayer which is not directly measurable. By analyzing the
experimental~\cite{Kasumov,Kasumov_priv} values $T_c =0.4$~K and $H_c =0.1$~T,
we get $\r\approx 111$ and $\r\approx 121$, respectively. Within the
experimental
accuracy of the bilayer's parameters, the two estimates for $\r$ should be
considered close. Interestingly, the value $\r\approx 121$ extracted from the
measured value of $H_c$ corresponds to the extremely narrow region of the
steep drop in $H_c (\r)$ (see Fig.~\ref{fig:h_c}).

Finally, we wish to remark on a peculiarity of real systems which can be
relevant when one compares our findings with an experiment. The point is that
during the fabrication of a bilayer, the interface between {\it S} and {\it N}
materials cannot be made ideally uniform. In other words, the local interface
resistance possesses spatial fluctuations. At the same time, as we have shown,
the bilayer's properties are highly sensitive to the interface quality, which
could lead to complicated behavior not reducing to the simple averaging of the
interface resistance embodied in $\r$. One possibility could be a
percolation-like proximity effect. We leave the study of inhomogeneity effects
for further investigation.

% 10 10 10 10 10 10 10 10 10 10 10 10 10 10 10 10 10 10 10 10 10 10 10 10 10
\section{Conclusions} \label{sec:conclusion}

We have studied, both analytically and numerically, the proximity effect in a
thin {\it SN} bilayer in the dirty limit. The layers were supposed to be thin
enough to ensure uniform properties of each layer across its thickness. The
strength of the proximity effect is governed by $\r$, the resistance of the
{\it SN} interface per channel.

The quantities calculated were $\D$, the order parameter; $n_S$, the density
of the superconducting electrons in the {\it S} layer; $T_c$, the critical
temperature; $E_g$ and $\nu(E)$, the minigap in the density of states and the
DOS itself; $H_c$ and $H_{c2}$, the critical magnetic field parallel to the
bilayer and the upper critical field perpendicular to the bilayer.

These quantities were calculated numerically over the entire range of $\r$.
For this purpose, the characteristics of the bilayer were assumed to be the
same as in the experiment by Kasumov {\it et al.}~\cite{Kasumov} that
originally stimulated our research (Ta/Au bilayer, $d_S/d_N =1/20$). In the
limit of an opaque interface, $\D$, $n_S$, $T_c$, $H_c$, and $H_{c2}$ approach
their BCS values. At the same time, $E_g$ does not coincide with the order
parameter $\D$, and $E_g\to 0$ when $\r\to\infty$, although in general, the
energy dependence of the DOS in the {\it S} and {\it N} layers, $\nu_S (E)$
and $\nu_N (E)$, approaches the BCS and normal-metal results, respectively.

The minigap $E_g$ demonstrates nonmonotonic behavior as a function of $\r$.
Analytical results for the two limiting cases of small and large $\r$ show
that in the Anderson limit, $E_g$ increases with increasing $\r$, whereas in
the limit of an opaque interface, $E_g$ tends to zero. Thus, $E_g$ reaches its
maximum in the region of intermediate $\r$.

The most interesting case of relatively low interface resistance (the Anderson
limit) has been considered analytically. The simple BCS relations between
$\D$, $n_S$, $T_c$, $H_c$, $H_{c2}$ are substituted by similar ones with $E_g$
standing instead of $\D$. The relation between the minigap $E_g$ and the order
parameter $\D$ in this limit is expressed by Eq.~(\ref{small}), implying that
in the case where $\tau_S < \tau_N$, the BCS relations are strongly violated
(by more than the order of magnitude for the above-mentioned Ta/Au bilayer).
The DOS in the {\it S} and {\it N} layers coincide, showing BCS-like behavior
with the standard peculiarity at $E=E_g$. It should be emphasized that
absolute values of $\r$ corresponding to the Anderson limit can be large; for
the Ta/Au bilayer this limit is already valid at $\r <80$.

All the results (except $H_c$) obtained for the bilayer also apply to more
complicated structures such as {\it SNS} and {\it NSN} trilayers, {\it SNINS}
and {\it NSISN} systems, and {\it SN} superlattices.

% -----------------------   ACKNOWLEDGMENTS   -------------------------------
\section*{Acknowledgments}

This research was supported by the RFBR grant \#~98-02-16252.

% -------------------------   REFERENCES   ----------------------------------

\end{multicols}


\begin{references}
\bibitem{Gamma}
The quantities $\Gamma_S$ and $\Gamma_N$ used by McMillan are analogous to
our quantities $1/\tau_S$ and $1/\tau_N$, respectively.
\bibitem{T_c_McMillan}
To avoid confusion, we note that Eq.~(39) from McMillan's
paper~\cite{McMillan} determining the critical temperature of the bilayer is
incorrect. The correct equation, following from Eqs.~(37), (38), (40) and
leading to Eq.~(41) of Ref.~[\onlinecite{McMillan}], reads $\ln (T_{CS}/T_C) =
(\Gamma_S/\Gamma)\, [ \psi ( \frac 12 +\Gamma/2\pi T_C ) -\psi (\frac 12 )]$.
\bibitem{free_el}
Certainly, the free-electron model does not describe the details of the
electronic structure of tantalum. However, this simplified model is used in
the very derivation of the Usadel equation. Also, we do not expect drastic
dependence of our results on the Fermi characteristics. To check this, we have
reproduced all the calculations with slightly different (within 10-15\%)
values of the Fermi energies, and found that the changes in the results amount
to, roughly speaking, a rescaling of the interface resistance $\r$ within
10\%.
\end{references}
\end{document}